\documentclass[useAMS,usenatbib]{mn2e}
\usepackage{graphicx}
\usepackage{color}

\newcommand{\ud}{\,\mathrm{d}}

\title[Galaxy properties in large scale environments]
	{Galaxy And Mass Assembly (GAMA): Trends in galaxy colours, morphology, and stellar populations with large scale structure, group, and pair environments}
\author[M. Alpaslan et al.]
	     {Mehmet Alpaslan$^{1,2,3}$, Simon Driver$^{2,3}$, Aaron S.G.~Robotham$^3$, Danail Obreschkow$^3$, \newauthor
	     Ellen Andrae$^4$, Michelle Cluver$^5$, Lee S. Kelvin$^6$, Rebecca Lange$^3$, Matt Owers$^{7,8}$, \newauthor
	     Edward N. Taylor$^9$, Stephen K. Andrews$^3$, Steven Bamford$^{10}$, Joss Bland-Hawthorn$^{11}$, \newauthor 
	     Sarah Brough$^{8}$, Michael J. I. Brown$^{12}$, Matthew Colless$^{13}$, Luke J. M. Davies$^3$, \newauthor
	     Elizabeth Eardley$^{14}$, Meiert W. Grootes$^4$, Andrew M. Hopkins$^{8}$, Rebecca Kennedy$^{10}$, \newauthor
	     Jochen Liske$^{15}$, Maritza A. Lara-L\'{o}pez$^{16}$, \'Angel R. L\'opez-S\'anchez$^{7,8}$, \newauthor
	     Jon Loveday$^{17}$, Barry F. Madore$^{18}$, Smriti Mahajan$^{19}$, Martin Meyer$^3$, \newauthor
	     Amanda Moffett$^3$, Peder Norberg$^{20}$, Samantha Penny$^{21}$, Kevin A. Pimbblet$^{12,22}$, \newauthor
	     Cristina C. Popescu$^{23,24}$, Mark Seibert$^{18}$, Richard Tuffs$^4$ \\
	     \\
	     Affiliations listed at the end of the paper.}

\pagerange{\pageref{firstpage}--\pageref{lastpage}}\pubyear{2015}

\begin{document}

\label{firstpage}

\maketitle

\begin{abstract}
We explore trends in galaxy properties with Mpc-scale structures using catalogues of environment and large scale structure from the Galaxy And Mass Assembly (GAMA) survey. Existing GAMA catalogues of large scale structure, group and pair membership allow us to construct galaxy stellar mass functions for different environmental types. To avoid simply extracting the known underlying correlations between galaxy properties and stellar mass, we create a mass matched sample of galaxies with stellar masses between $9.5 \leq \log{M_*/h^{-2} M_{\odot}} \leq 11$ for each environmental population. Using these samples, we show that mass normalised galaxies in different large scale environments have similar energy outputs, $u-r$ colours, luminosities, and morphologies. Extending our analysis to group and pair environments, we show galaxies that are not in groups or pairs exhibit similar characteristics to each other regardless of broader environment. For our mass controlled sample, we fail to see a strong dependence of S\'{e}rsic index or galaxy luminosity on halo mass, but do find that it correlates very strongly with colour. Repeating our analysis for galaxies that have not been mass controlled introduces and amplifies trends in the properties of galaxies in pairs, groups, and large scale structure, indicating that stellar mass is the most important predictor of the galaxy properties we examine, as opposed to environmental classifications.
\end{abstract}

\begin{keywords}
 galaxies: evolution -- galaxies: clusters: general -- galaxies: luminosity function, mass function -- galaxies: stellar content -- large-scale structure of Universe
\end{keywords}

\section{Introduction}

No galaxy is an island. The close surroundings of a galaxy (i.e. within a few tens of Mpc) play an important role in shaping its evolution (e.g. \citealp{Avila-Reese2005,Blanton2005,Hahn2007,Fakhouri2009}). Stellar populations are particularly sensitive to environment: the proximity of nearby or merging galaxies has been shown to trigger gas collapse within galaxies, which leads to an increased rate of star formation \citep{Toomre1972,Quinn1993,Lewis2002,Gomez2003,Porter2008}. Other observational signatures of environmental effects include differences in colour \citep{Kreckel2012}, stellar mass \citep{Chabrier2003}, morphology \citep{Dressler1980,Butcher1984,Dressler1997}, the luminosity function \citep{Croton2005,McNaught-Roberts2014}, and disrupting the gas content of a galaxy \citep{Beygu2013a,Benitez-Llambay2013}.

Beyond the close surroundings of a galaxy, there is also ample evidence to show that, for galaxies in a cluster, the cluster environment has some influence on its properties. Galaxies do appear to be more red in denser environments (e.g. \citealp{Baldry2006,Bamford2009,Peng2010}). On the other hand, star formation appears to be suppressed in galaxies that are infalling into clusters \citep{Lewis2002,Balogh2007}; though more recent work by \citet{Brough2013} fails to detect significant trends in star formation rates with environment. Similarly, recent work by \citet{Wijesinghe2012} determines that star formation rates of morphologically classified galaxies depend, to first order, on stellar mass and not environment. Another well known phenomenon is the tendency for galaxies in different group environments to have distinct luminosity functions (e.g. \citealp{Norberg2002,DePropris2003,Croton2005,Zehavi2005,Robotham2006,Robotham2010,Zehavi2011,Masaki2013,McNaught-Roberts2014,Eardley2014}).

Galaxies, however, do not just live amongst their immediate neighbours; instead, they are embedded within the complex large scale structure of the Universe. This structure is characterised by long, linear filaments of galaxies that stretch between highly dense clusters containing a vast number of galaxies. In between filaments and clusters lie voids, which are largely empty regions of space that contain a mere handful of galaxies per unit volume. Does this so-called `Cosmic Web' have any noticeable impact on galaxies? In other words, is a galaxy in a filament distinctly and systematically different to that inside a void?

Research in understanding the Cosmic Web is actively ongoing. A number of algorithms and methods exist to detect and classify large scale structure (e.g., \citealp{Doroshkevich2004,Pimbblet2005a,Colberg2007,Forero-Romero2009,Murphy2011,Smith2012a,Cautun2012,Alpaslan2013a,Eardley2014}). These algorithms primarily focus on using positional (and where possible, velocity) information in the distribution of galaxies to trace and identify not only filamentary structures, but also voids and clusters (often referred to as `knots'). These sophisticated algorithms, together with current generation redshift surveys make it possible to compare and correlate the properties of galaxies found in a range of environments.

The Galaxy and Mass Assembly (GAMA)\footnote[1]{http://www.gama-survey.org/} survey (\citealp{Driver2009,Driver2011}, Liske et al. 2014) is a joint European-Australian project which combines multiwavelength photometric data from a large number of ground and space-based programs with a highly complete spectroscopic campaign conducted using the AAOmega spectrograph on the Anglo-Australian Telescope at Siding Spring Observatory, Australia. At the time of writing, GAMA has access to data in the following bands (see Driver et al. in prep): Galaxy Evolution Explorer (GALEX; \emph{FUV, NUV}), Sloan Digital Sky Survey (SDSS; \emph{ugriz}), UKIRT Infrared Deep Sky Survey  (UKIDSS; \emph{YJHK}),  VISTA Kilo-degree Infrared Galaxy survey (VIKING; \emph{ZYJHK}), Wide-Field Infrared Survey Explorer (WISE; \emph{W1-4; MIR}), and Herschel ATLAS (H-ATLAS; \emph{PACS, SPIRE; FIR}) and Giant Metrewave Radio Telescope and shortly Australian SKA Pathfinder (radio). The spectroscopic component of the survey contains $\sim$250,000 spectra for galaxies out to $r = 19.8$ mag across 5 fields: centred at $\alpha$ = 9h, $\delta$ = 0.5 deg (G09), $\alpha$ = 12h, $\delta$ = -0.5 deg (G12) and $\alpha$ = 14h30m, $\delta$ = 0.5 deg (G15), $\alpha$ = 2h, $\delta$ = -8.125 deg (G02) and $\alpha$ = 23h and $\delta$ = -32.5 deg (G23). The three equatorial fields (G09, G12 and G15) are $12 \times 5$ degrees each, and the two southern fields (G02 and G23) are respectively $8.6 \times 2.5$ and $12 \times 5$ degrees.

One of the principal advantages of GAMA for studies of intergalactic structure is the high spectroscopic target density; which averages to 1,050 galaxies per square degree, and a spectroscopic completeness of 98.42\% in the three equatorial fields, with associated velocity uncertainties of $\sigma_v \approx 50$ kms$^{-1}$ (Liske et al. 2014). This is achieved via an observing strategy which revisits the same patch of sky, on average, 10.3 to 10.9 times (Liske et al. in prep), with different fibre configurations on the 2dF instrument, in order to observe every possible target \citep{Robotham2010a}. These highly complete and dense data have led to the creation of the GAMA Galaxy Group Catalogue (\citealp{Robotham2011}, G$^3$C) and the GAMA Large Scale Structure Catalogue (\citealp{Alpaslan2013a}, GLSSC), both of which will be discussed in greater detail in Section \ref{sec:data}. These data sets allow us to tackle outstanding questions in galaxy evolution from an empirical perspective, and will help in placing robust constraints on how the large scale structure of the Universe affects galaxy evolution; particularly morphology, colour, and stellar populations. Ultimately a detailed study of the gas content will also be required, possible via projects such as the ASKAP DINGO survey \citep{Meyer2009}.

In this work, we seek to make some initial advances into understanding to what extent (if at all) large scale structure has an impact on the evolution of galaxies. Given the broadness of this topic, we choose to focus on a few, relatively well understood properties of galaxies: colour, brightness, morphological properties (S\'{e}rsic index, ellipticity, effective radius, and visually identified structure), gas metallicity, and total spectral energy density. For each of these properties, we seek to find any possible trends related to large scale structure. In other words, do galaxies that exist in different types of large scale structure inhabit different parts of the parameter space of galaxy properties? How do these trends compare to those of galaxies in groups, or pairs?

In Section \ref{sec:data} we give further detail on the G$^3$C and the GLSSC, as well as other GAMA catalogues of the various galaxy properties that were used in this investigation. In Section 3, we display galaxy stellar mass functions for GAMA galaxies in different environments. Section \ref{sec:results} begins by defining a mass normalised sample of galaxies, which we then use to look for trends in galaxy properties within different types of environment. Finally, in Section 5 we briefly examine consequences of our mass normalisation process and repeat our analysis for different mass ranges. In Section 6 we discuss these results and summarise them. Throughout this paper, consistent with the cosmology used in \citet{Robotham2011,Merson2012,Alpaslan2013a}, we adopt $\Omega_{\mathrm{m}} = 0.25,\; \Omega_{\Lambda}=0.75,\; H_0 = h\; 100 \mathrm{\;km\; s}^{-1}\;\mathrm{Mpc}^{-1}$. All of the analysis in this work has been done using the R programming language \citep{Rproj}.

\section{Data}
\label{sec:data}

The large scale catalogue we use is the GLSSC (\citealp{Alpaslan2013a}, \texttt{FilamentFindingv01}), an observationally motivated catalogue of large scale structure. It classifies galaxies in the three equatorial GAMA fields as belonging to filaments or voids. The catalogue also introduces a third, interstitial structure of galaxies dubbed `tendrils,' which typically consist of delicate (but coherent) strings of 6--10 galaxies that emerge from filaments and terminate either in voids, or in other filaments. Tendrils are discussed in greater detail in \citet{Alpaslan2014b}. The catalogue contains 45,542 galaxies with $M_r \leq -19.77 + 5\log h$ mag across all three equatorial fields, each assigned to a filament, tendril, or void. The full sample of galaxies is volume limited to $M_r \leq -19.77 + 5\log h$ mag and extends out to $z = 0.213$. Considering the GAMA selection cut of $r = 19.8$ mag, this redshift and its associated absolute magnitude cut are chosen such that they maximise the number of galaxies in the GLSSC. The catalogue also identifies 643 filaments, composed of an average of 8 groups and spanning up to 100 h$^{-1}$ Mpc. The appendix of \citet{Alpaslan2013a} goes into further detail on the actual composition of the catalogue. 

At its heart, the algorithm that generates the GLSSC relies on a slightly modified \emph{minimal spanning tree} (MST) algorithm, similar to that of \citet{Doroshkevich2004} and \citet{Colberg2007} and uses a multiple pass approach, similar to \citet{Murphy2011}. Initially, a volume limited sample of groups from the G$^3$C are used as nodes of a minimal spanning tree to identify filaments; galaxies within a distance $r$ of each filament are then associated with that filament. These galaxies, within and outside of groups, are together referred to as filament galaxies. The remaining galaxies (not in filaments) are processed through a second minimal spanning tree to identify tendrils and tendril galaxies; and any galaxy beyond a distance $q$ from a tendril is considered to be a very isolated galaxy, which we call a void galaxy. Note that this approach makes no attempt to physically identify voids. The values for $r$ and $q$ are chosen such that the two point correlation function $\xi_2 (r)$ of void galaxies is minimised i.e. our fiducial assumption is that voids should have a minimal amount of coherent structure due to the low net accelerations experienced over the lifetime of the Universe. A third parameter, $b$ represents the maximum allowed distance between groups in order for them to be considered to be in the same filament; this is chosen such that at least 90\% of groups with $L_* \geq 10^{11} L_{\odot}$ are located in filaments. The GLSSC is generated with $b = 5.75 h^{-1}$ Mpc; $r = 4.12 h^{-1}$ Mpc; and $q = 4.56 h^{-1}$ Mpc.

\subsection{Additional GAMA catalogues}

Here we briefly introduce each of the other catalogues that are used for the analysis in this work. Where possible, we reference the publication that accompanies the catalogue, and suggest that the more curious reader refers to these for additional detail. Where no reference is given, the catalogue has been provided by one of the authors of this paper. We consider only data for galaxies in the three equatorial GAMA fields.

\subsubsection{Galaxy spectra}

This catalogue (Liske et al. in prep, \texttt{SpecCatv25}), contains spectra for all galaxies observed using the 2dF instrument at the Anglo-Australian Telescope in Australia, for GAMA. Spectra obtained at the AAT were redshifted by observers at the telescope, as well as via an automated algorithm \textsc{Autoz} (see \citealp{Baldry2014} for details on \textsc{Autoz}, and Liske et al. in prep for details on the spectroscopic campaign). On average, redshift measurements for the spectra have an associated error $\sigma$ of approximately 27 km s$^{-1}$; and only 0.2\% of redshifts are considered to be incorrect (\citealp{Baldry2014}, Liske et al. in prep). Some spectra from previous surveys (e.g. SDSS) are also included in this catalogue. These spectra (and the redshifts derived from them) form the most important data asset for the catalogues of environment discussed below.

\subsubsection{The GAMA Group Catalogue (G$^3$C)}
\label{sec:g3c}

The G$^3$C (\citealp{Robotham2011}, \texttt{GroupFindingv07}) is a group catalogue that has been put together using a slightly modified friends-of-friends algorithm that considers galaxies to be in a group if they are grouped both along the line of sight, as well as when projected onto the sky. This successfully accounts for redshift space distorsions caused by the peculiar velocities of galaxies in groups. The catalogue contains 23,838 groups with 2 or more members out to $r = 19.8$ mag; these groups contain approximately 40\% of all galaxies in GAMA. This catalogue also provides galaxy pairing information, which is also used in this work and studied in \citet{Robotham2014}.  Parameters for the groupfinding algorithm is calibrated by being run on a series of GAMA mock galaxy catalogues, described in \citet{Merson2012}, and the optimal linking lengths have been verified as theoretically optimal in recent work by \citet{Duarte2014}. The mock catalogues are designed to mimic the geometry of the fields observed by GAMA, as well as replicate the galaxy luminosity function of the survey. They are assembled by populating haloes taken from the Millennium Simulation \citep{Springel2005} using the \citet{Bower2006} \textsc{Galform} semi-analytic galaxy formation model.

\subsubsection{Aperture matched photometry}

Aperture matched photometry for GAMA (Liske et al. in prep, \texttt{ApMatchedCatv05}) is measured by running Source Extractor \citep{Bertin1996,BertinE.2011} on imaging data from the SDSS and UKIDSS (for $ugriz$ bands and $YJHK$ bands respectively) in matched aperture mode. This allows us to place apertures on sources in the $r$-band (as this has the highest imaging quality and defines the main survey spectroscopic sample selection) and measure fluxes in all bands. The imaging data have been preprocessed into a series of 27 (9 bands $\times$ 3 fields) large mosaics using SWarp \citep{Bertin2002} and have been normalised to a common zeropoint and convolved to a common PSF of 2 arcseconds (see \citealp{Hill2011} and Driver at al. in prep).
\subsubsection{Stellar masses}
\label{sec:masscatchat}

The GAMA stellar masses (\citealp{Taylor2011}, \texttt{StellarMassesv16}) are estimated using photometry in the restframe $ugriz$ bands from the aperture matched catalogue \texttt{ApMatchedCatv05}, to which a series of synthetic spectra are fit. The spectra are designed to incorporate an exponentially decaying star formation history, using \citet{Bruzual2003a} models with a \citet{Chabrier2003} IMF and the \citet{Calzetti2000} dust obscuration law. The stellar masses are determined by integrals that are weighted to the probability of each fit. This catalogue also provides dust and extinction corrected apparent and absolute magnitudes, as well as restframe and extinction (Galactic and self-attenuation) corrected $u-r$ and $g-i$ colours, which are also used in this work. The extinction correctios are estimated during the sythentic spectral fits and have associated uncertainties of $A_v \approx 0.15$ mag \citep{Taylor2014}, but this is likely to be an overestimation as the random errors on the rest-frame colours are on the order of 0.05 mag. As explained in \citet{Taylor2014}, the amount of dust from which $A_v$ is derived for each galaxy cannot be negative, leading to a possible overstimation of $A_v$ for dust-poor galaxies. This will translate into an error in $(g-i)$ colours of up to 0.1 mag; however, these errors have no discernible impact on our results.

\subsubsection{S\'{e}rsic photometry and morphology}
This catalogue provides single component S\'{e}rsic fits of galaxies, computed (\texttt{SersicPhotometryv07}) using the SIGMA software which wraps around various astronomy packages including GALFIT3 \citep{Kelvin2012,Peng2011}. Along with providing a S\'{e}rsic index $n_r$ for each galaxy, this catalogue also provides some other morphological parameters such as effective radius $R_e$ and ellipticity, defined as $1 - b/a$, where $a$ is the ratio of the semi-major and $b$ the minor axes of the galaxy. An ellipticity value of 0 being given to galaxies whose semi-major and semi-minor axes are equal (i.e. round galaxies).

\subsubsection{Emission line measurements and metallicities}

This catalogue measures fluxes and equivalent widths for spectra for all GAMA galaxies. Measurements are made for 11 different emission features, including H$\alpha$, H$\beta$, O\textsc{iii} $\lambda5007$, and N\textsc{ii} $\lambda6583$, using single and double Gaussian fits to continuum-substracted spectra. This is done in a manner similar to that described in \citet{Gunawardhana2013}. Gas metallicities of galaxies are then calculated using the O3N2 index. This is defined as:

\begin{equation}
	\mathrm{O3N2} = \log_{10} \left(\frac{[\mathrm{O\textsc{iii}}]\, \lambda5007 / \mathrm{H}\beta}{[\mathrm{N\textsc{ii}}]\, \lambda6583 / \mathrm{H}\alpha} \right)
\end{equation}

\noindent from which it is possible to calculate the O/H metallicity indicator as $[12+\log (\mathrm{O/H})] = 8.73 - 0.32 \times \mathrm{O3N2}$ as per the prescription of \citet{Pettini2004}. Line strengths are emission measurements and have been corrected for dust attenuation, and have been calibrated to match metallicities given by \citet{Tremonti2004}, as per the prescription given in \citet{Lara-Lopez2013}.

\subsubsection{GALEX photometry}
This catalogue provides GALEX NUV and FUV photometry for the GAMA II equatorial survey regions. The GALEX ultraviolet catalogue (\texttt{GalexPhotometryv02}; Andrae et al. in prep) is a combination of archival data and pointed observations on equatorial GAMA fields. The archival data have been used to extend the ultraviolet coverage of the GAMA regions as much as possible.

\subsubsection{WISE photometry}
Photometry for the four mid-infrared WISE bands is provided in \texttt{WisePhotometryv01} for GAMA galaxies in all three equatorial fields that are matched to WISE observations. The catalogue does not include sources that have no GAMA match within 3 arcseconds of a WISE source (see \citealp{Cluver2014} for details). The photometry for each source is from the All-Sky WISE Data Release with standard aperture measurements for unresolved sources and isophotal photometry for resolved sources. Profile-fit measurements are also provided.

\subsubsection{Multi-band photometry}
The process of amalgamating data from multiple sources into a single photometry catalogue (\texttt{20BandPhotomv02}) is described in Driver et al. (in prep) and builds on the work described in \citet{Hill2011}; here we provide a brief summary. The GALEX (\emph{FUV, NUV}), aperture matched photometry (\emph{u-K}) and WISE (\emph{W1-4}) catalogues are combined to PACS and SPIRE photometry derived from the PACS/SPIRE maps by measuring the flux in the appropriate optically defined aperture convolved with the appropriate PACS/SPIRE PSF\footnote{Where objects overlap care is taken to divide the flux between the two objects following the prescription outlined in Appendix A of \citet{Bourne2012} (using the aperture matched photometry catalogue as the input catalogue to deblend data), see also Driver et al. in prep.}. This combination is done by exact name matching to unique identifiers that are given to each matching galaxy using \textsc{topcat} \citep{Taylor2005}. The combined catalogue is corrected for Galactic extintion using $E(B-V)$ values provided by \citet{Schlegel1998a} and the coefficients listed in Liske et al. (in prep). The final product is a catalogue that contains invidual flux measurements in Janksys across all bands for each individual GAMA galaxy, with dummy flux values included if that galaxy has not been surveyed in a particular band. This data is used only in Section \ref{sec:csed}.

\subsubsection{Visual morphology}
For a sub-sample of galaxies with $0.002 \leq z \leq 0.1$, $r < 19.8$ mag, this catalogue (\texttt{VisualMorphologyv02}) provides visual classifications of galaxy morphology as being either elliptical or not-elliptical. The classification was performed on postage stamps generated from three colour giH band images with arctan scaling from the SDSS and VIKING or UKIDSS LAS data. The postage stamp sizes were set to a constant value of 30 kpc x 30 kpc, except if the implied size is greater than 100 pixels, in which case it was set to that value.

\subsubsection{Surface density}
The nearest neighbour surface density, $\Sigma_5$ (\citealp{Brough2013}, EnvironmentMeasuresv5), is calculated for all GAMA galaxies with reliable redshifts (nQ $> 2$; \citealp{Driver2011}). The surface density is defined using the projected comoving distance to the fifth nearest neighbour (d5) with $\pm$1000 km s$^{−1}$ within a volume-limited density-defining population: $\Sigma_5=5/\pi d^2_5$. The density-defining population has absolute SDSS petrosian magnitudes $M_r < M_{r, limit} − Qz$, k-corrected to $z = 0$ following \citet{Loveday2011}, where $M_{r, limit}= −20.0$ mag and $Q$ defines the expected evolution of $M_r$ as a function of redshift ($Q = 0.78$; Loveday et al. in prep). Densities are then corrected for the survey r-band redshift completeness as $\Sigma_5 = \Sigma_{5, raw} \times 1$/completeness. Galaxies where the nearest survey edge is closer than the fifth nearest neighbour are flagged as upper limits.

\subsection{Sample selection}

\begin{figure}
	\centering
	\includegraphics[width=0.5\textwidth]{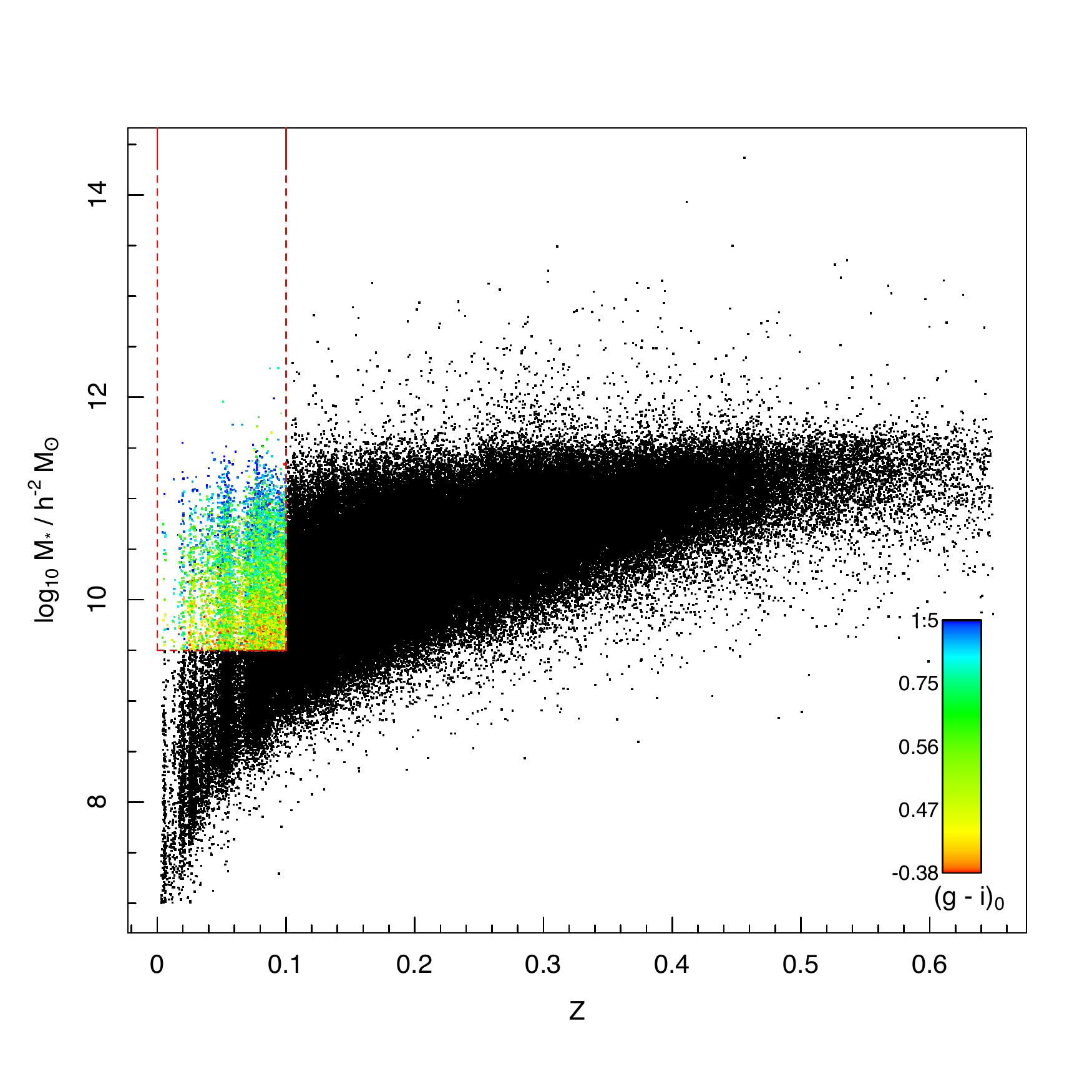}
	\caption{Stellar mass as a function of redshift for all GAMA galaxies. Galaxies with $z \leq 0.1$ and $M_* \geq 10^{9.5} h^{-2} M_{\odot}$ are coloured by their restframe $g-i$ colour. Redshift and mass limits are shown as red lines. These limits are chosen as they are shared by all the catalogues we consider in this work.}
	\label{fig:sample}
\end{figure}

Galaxy properties are well known to correlate with stellar mass. To ensure we recover variations due to environment rather than these known stellar mass dependencies, it is vital that we first remove the mass dependency from our sample. We do this by forcing our samples to be mass complete, using stellar mass data from the \texttt{StellarMassesv16} catalogue. We choose the limits $z \leq 0.1$ and $M_* \geq 10^{9.5} h^{-2} M_{\odot}$; such that we can use the maximal number of galaxies from all catalogues considered in this work, each with its own redshift limits. Our sample is shown in Figure \ref{fig:sample}, where it is highlighted according to each galaxy's restframe $g - i$ colour, and the red lines mark our limits. In \citet{Alpaslan2013a} it is established that the GLSSC is mass complete only to $\log M_*/h^{-2} M_{\odot} \geq 10.61$ (but does include galaxies of lower mass than this). To account for the discrepancy between this mass limit and our chosen sample, we take all galaxies with $\log M_*/h^{-2} M_{\odot} \geq 9.5$ with $z \leq 0.1$, and $M_r \leq -19.77 + 5\log h$ mag and classify them as belonging to the same type of large scale structure (i.e. filament, tendril, or void) as their nearest neighbour using the GLSSC. Our low mass large scale structure sample contains 7195 volume limited galaxies.

In \citet{Alpaslan2014b} we show that void galaxies in the GLSSC are isolated and not associated with any structures by making use of the line correlation function. Introduced in \citet{Obreschkow2013}, the line correlation function $l(r)$ measures the amount of linearity in structures of length $r$ in a way that is analogous to how the two-point correlation function measures clustering on length scales of $r$. The line correlation function takes the phase factor information of the galaxy density field (i.e. $\hat\epsilon(\vec{k})\equiv\hat{\delta}(\vec{k})/|\hat{\delta}(\vec{k})|$, where $\hat{\delta}(\vec{k})$ is the Fourier transform (FT) of a galaxy density field $\delta(\vec{r})$) and calculates the normalised three-point correlation of the inverse FT of $\hat\epsilon(\vec{k})$ for three points on a straight line separated by a distance $r$. For an illustration of how the line correlation function can distinguish between random, spherical, and linear overdensities, we refer the reader to Figure 3 in \citet{Obreschkow2013}.

We repeat this calculation for the galaxies in our low-mass GLSSC sample in order to confirm that the process of nearest galaxy association described above has not resulted in the association of galaxies in structures with void galaxies. As was done in \citet{Alpaslan2014b}, we measure the filamentarity of each galaxy population by calculating its so-called excess line correlation with respect to a random points set. The excess line correlation is defined as $\Delta l(r)=[l(r)-l_0(r)]\sqrt{f}$, where $l(r)$ is the line correlation of galaxies in the GLSSC placed into a box of side 150 h$^{-1}$ Mpc, $l_0(r)$ is the line correlation of an equal number of galaxies placed randomly into a cone identical to the GAMA geometry within this box, and $f$ is the volume fraction of the GAMA fields within this box. Including $f$ ensures that $\Delta l(r)$ is approximately independent of the volume of the box (see Section 3.4 in \citealp{Obreschkow2013}).

\begin{figure}
	\centering
	\includegraphics[width=0.5\textwidth]{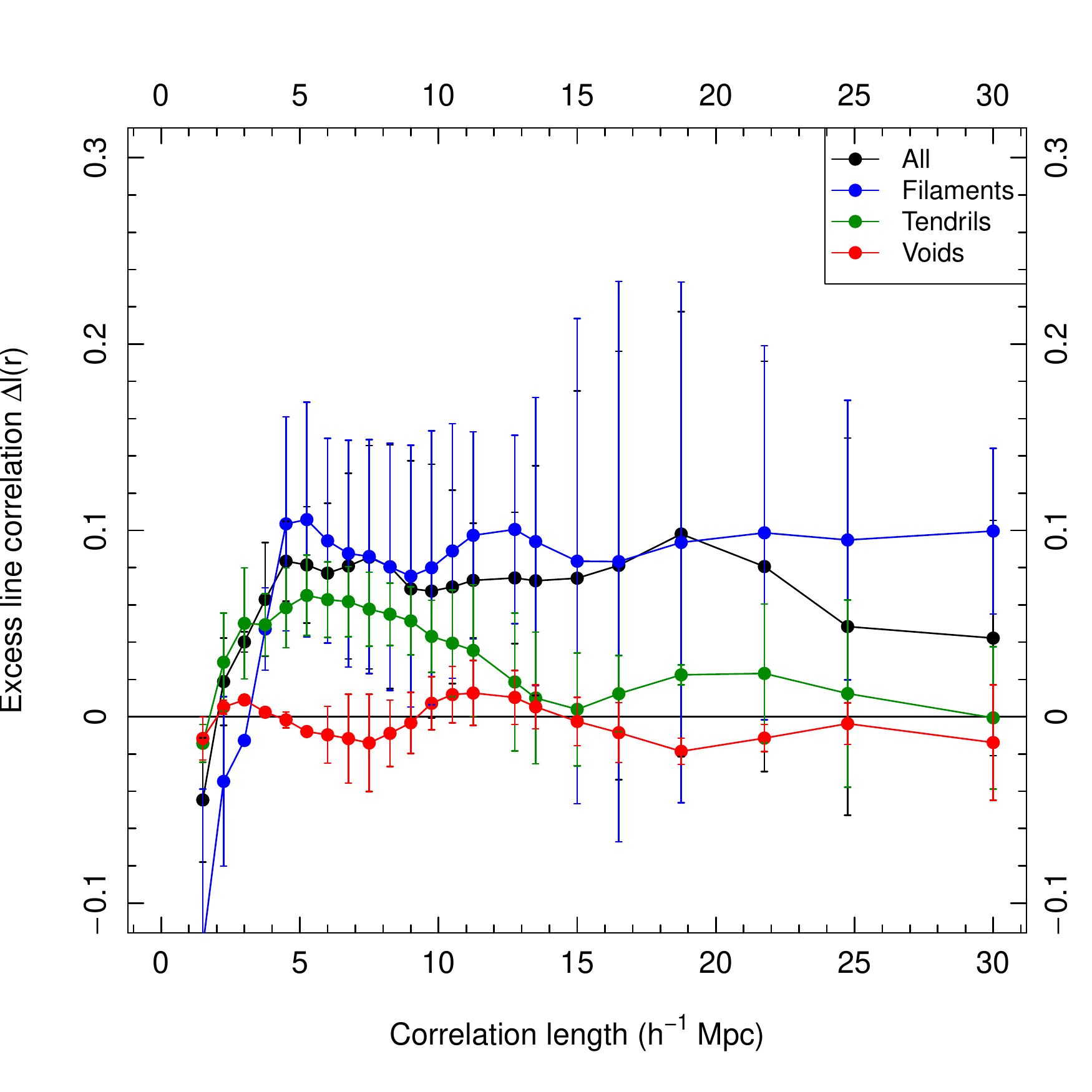}
	\caption{The excess line correlation $\Delta l(r)$ of galaxies in filaments, tendrils, voids, and all galaxies shown in blue, green, red, and black respectively. We compute $\Delta l(r)$ individually for each population of galaxies in all three GAMA fields, and show the averaged results in this figure. The errors shown are the standard deviation in the excess line correlation across the three GAMA fields. Most notably, we still retain the 0 signal for void galaxies across large distances. Readers making a direct comparison between this figure and Figure 2 in \citet{Alpaslan2014b} must note that the line correlation calculation method has since been updated (see Wolstenhulme et al., submitted for details). This is mainly reflected by a change in the vertical normalization of $\Delta l(r)$.}
	\label{fig:linecorr}
\end{figure}

In Figure \ref{fig:linecorr} we show the excess line correlation of galaxies in the low mass GLSSC sample averaged across all three equatorial GAMA fields. Notably, filament galaxies continue to have the highest $\Delta l(r)$ signal, while void galaxies show no filamentarity. Errors shown are the variance of $\Delta l(r)$ across all three GAMA fields, with the scatter due to the smaller number of points in this sample. The lack of filamentarity in the void galaxy sample confirms that we have not inadvertently associated galaxies within structures with the void galaxy population. Figure \ref{fig:linecorr} confirms the linearity (or lack thereof) present in the three large scale structure classifications.

We further test our method by taking the GLSSC and randomly removing 50\% of the galaxies in each environment (filament, tendril, and void). We then apply the reclassification scheme described in the beginning of this section to these removed galaxies, and compare that classification to what the GLSSC classification is. This random rejection process is realised 100 times, and the average recovery rate with the correct classification is 90\% for all three categories of large scale structure (with a standard deviation sub-1\%, indicating that this process is robust to randomness). This high recovery rate is a strong indicator that the reclassification scheme does not create large biases in our large scale structure sample. The $\approx 10\%$ of galaxies that are misclassified are usually those whose neighbours were also randomly removed, resulting in the nearest structure being of another type. Their spatial distribution is randomised within each realisation, and the mean distance between a reclassified galaxy and its nearest neighbour is 1.97 $h^{-1}$ Mpc for this test case, and approximately 10\% greater for the reclassification scheme used to generate the low mass sample. 

\section{Galaxy stellar mass functions by environment}

\begin{figure}
	\centering
	\includegraphics[width=0.5\textwidth]{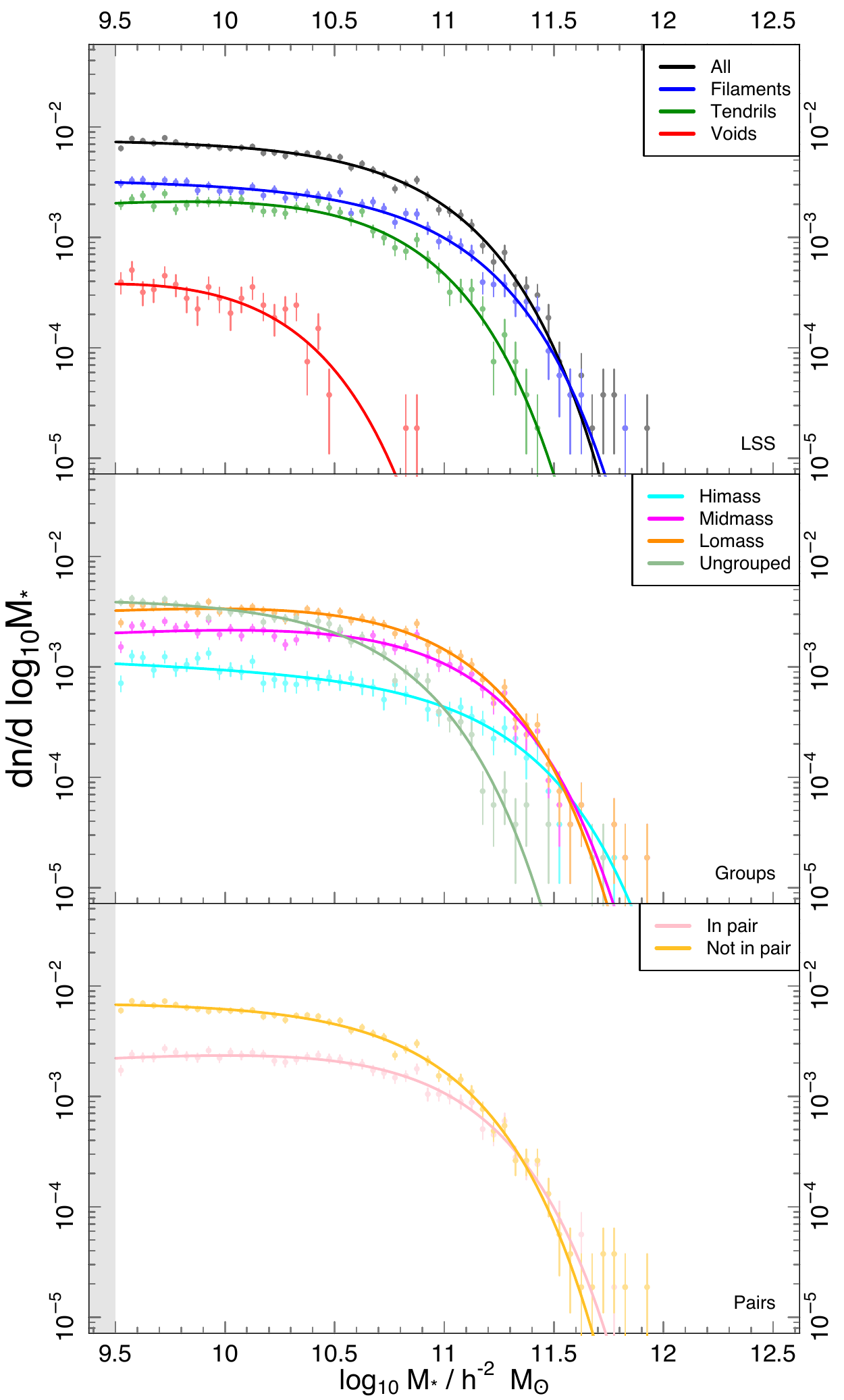}
	\caption{Galaxy stellar mass functions of galaxies in different environments by 0.1 dex bins in log stellar mass, with single-Schechter function fits to each. The region shaded in grey marks our mass selection limit. Each panel shows GSMFs for galaxies in different types of environment: large scale structure (top), groups (middle), and pairs (bottom), with the total GSMF for all galaxies shown as the black line in the top panel. Each set of GSMFs adds up to this total set. See Figure 3 in \citet{Kelvin2014a} for GSMFs of galaxies by their morphological type.}
	\label{fig:stellarMassFunc}
\end{figure}

In order to broaden our investigations beyond the simple filament, tendril and void sub-samples we also look to now incorporate some of the additional information available within the GAMA database, i.e., pairings and groups. Figure \ref{fig:stellarMassFunc} shows the galaxy stellar mass functions (GSMFs; see \citealp{Baldry2012,Kelvin2014a}) of galaxies in a variety of different environments, including those in the low mass GLSSC sample. The environmental types for which we calculate GSMFs in Figure \ref{fig:stellarMassFunc} are: filaments, tendrils, voids (as defined in the GLSSC), galaxies in groups with high mass halos, mid mass halos and low mass halos (defined as $M_H \geq 10^{14} M_{\odot}$, $10^{13} \leq M_H \leq 10^{14} M_{\odot}$ and $M_H \leq 10^{13} M_{\odot}$ respectively), galaxies not in groups, galaxies in pairs, and galaxies not in pairs (any two galaxies with a physical projected separation of 100 h$^{-1}$ kpc and velocity separation of 1000 kms$^{-1}$ are considered to be in a pair). Following the prescription of \citet{Kelvin2014a}, we calculate the GSMF in the form of a Schechter function: 

\begin{equation}
	\Phi (\widetilde{M}) \ud \widetilde{M} = \ln(10) \phi^* 10^{(\widetilde{M}-\widetilde{M}^*)(\alpha+1)} \exp \left(-10^{(\widetilde{M}-\widetilde{M}^*)}\right) \ud \widetilde{M}
\end{equation}

\noindent where $\widetilde{M} \equiv \log_{10} M$. The fit parameters for the stellar mass functions are given in Table \ref{table:schecfits}. The errors in the parameters are estimated via jackknife resampling, with $\sigma^2 = \frac{N-1}{N} \sum^N_{i=1} (x_j - x)^2$ where $x$ is the best fitting parameter and $x_j$ is the best fitting parameter as given from a jackknife resampled version of the data. The jackknifing is done in such a way as to split the data into $N=10$ equal sized bins. Across all populations, the slope at the low mass end $\alpha$ is statistically consistent with being constant. The rise at the low mass end of the overall GAMA GSMF \citep{Baldry2012} must therefore be caused by new low-mass populations being sampled, and not due to any fluctuations in the numbers of other populations below $M_*$.

\begin{table*}
	\footnotesize
	\centering
	 \caption{Parameters for the single Schechter function GSMF fits shown in Figure \ref{fig:stellarMassFunc}. The columns are the knee in the Schechter function ($M^*$), the slope at the low mass end ($\alpha$), the normalisation ($\phi*$; in units of dex$^{-1}$ Mpc$^{-3}$), and the goodness of fit $\chi^2/\nu$. The fractional integrated stellar mass from the fits is also shown, for subdivisions of environment (marked by the horizontal lines in the table). Errors are estimated from jackknife resampling.}
	\begin{tabular}{l|ccccccc}
	 \hline
	 &$\log M^* / M_{\odot}$&$\alpha$&$\phi^*/10^{-3}$&$\chi^2/\nu$&$\int{\Phi (M) \ud M}$\\
	 \hline
	 All&$10.82 \pm 0.02$&$-0.97 \pm 0.02$&$2.00 \pm 0.13 $&1.16&100\%\\
	 \hline
	 Filaments&$10.88 \pm 0.02$&$-0.99 \pm 0.02$&$1.52 \pm 0.07$&1.2&57\%\\
	 Tendrils&$10.66 \pm 0.03$&$-0.86 \pm 0.05$&$1.55 \pm 0.01$&1.09&39\%\\
	 Voids&$10.31 \pm 0.55$&$-1.24 \pm 0.62$&$0.15 \pm 0.19$&2.31&4\%\\
	 \hline
	 HiMass&$11.16 \pm 0.10$&$-1.06 \pm 0.07$&$0.27 \pm 0.06$&1.25&10\%\\
	 MidMass&$10.93 \pm 0.025$&$-0.87 \pm 0.02$&$1.12 \pm 0.05$&1.59&24\%\\
	 LoMass&$10.86 \pm 0.02$&$-0.87 \pm 0.02$&$1.74 \pm 0.06$&1.16&36\%\\
	 Ungrpd&$10.64 \pm 0.04$&$-1.07 \pm 0.03$&$1.35 \pm 0.11$&1.60&30\%\\
	 \hline
	 inPair&$10.89 \pm 0.02$&$-0.87 \pm 0.03$&$1.22 \pm 0.06$&1.28&30\%\\
	 notinPair&$10.80 \pm 0.02$&$-1.02 \pm 0.02$&$2.50 \pm 0.12$&1.01&70\%\\
	 \hline
	 \end{tabular}
	 \label{table:schecfits}
\end{table*}


\section{Are galaxy properties influenced by large scale structure?}
\label{sec:results}

\subsection{Mass normalisation}
\label{sec:massnorm}

From Figure \ref{fig:stellarMassFunc} it is evident that galaxies in different environments have vastly different stellar mass functions. Given that stellar mass most directly influences galaxy evolution (e.g. \citealp{Oesch2010,Robotham2013a}), performing our analysis on our galaxy sample as is is more likely to draw out trends in stellar mass; or conversely, blur out any trends caused by environment. 

\begin{figure}
	\includegraphics[width=0.5\textwidth]{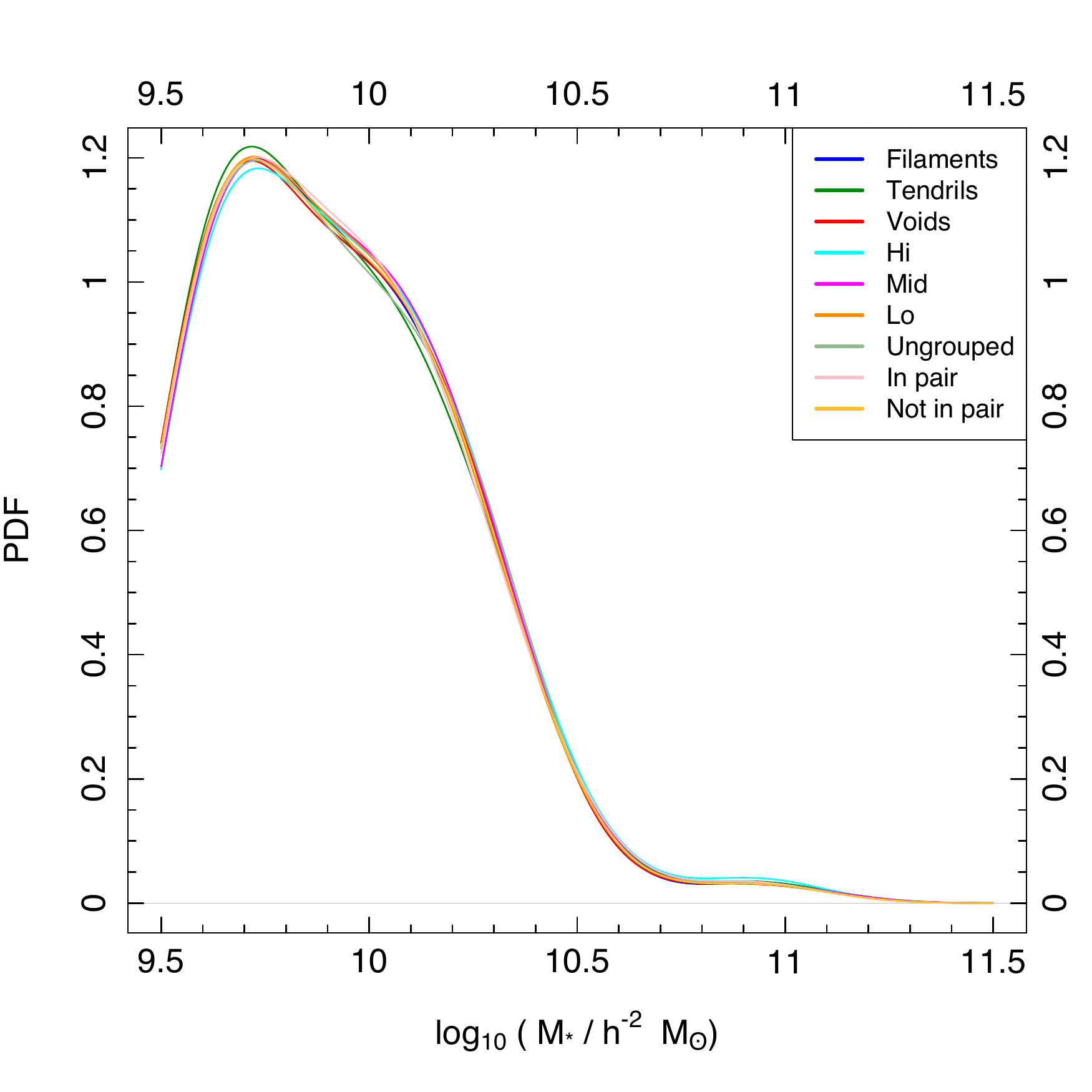}
	\caption{Kernel density estimates of the stellar mass distribution in mass normalised galaxy populations (colour coded in the same way as Figure \ref{fig:stellarMassFunc}). The bumps in density are due to Poisson noise in the void sample, which are then reproduced in the other populations by design.}
	\label{fig:massnorm}
\end{figure}
\begin{figure}
	\includegraphics[width=0.5\textwidth]{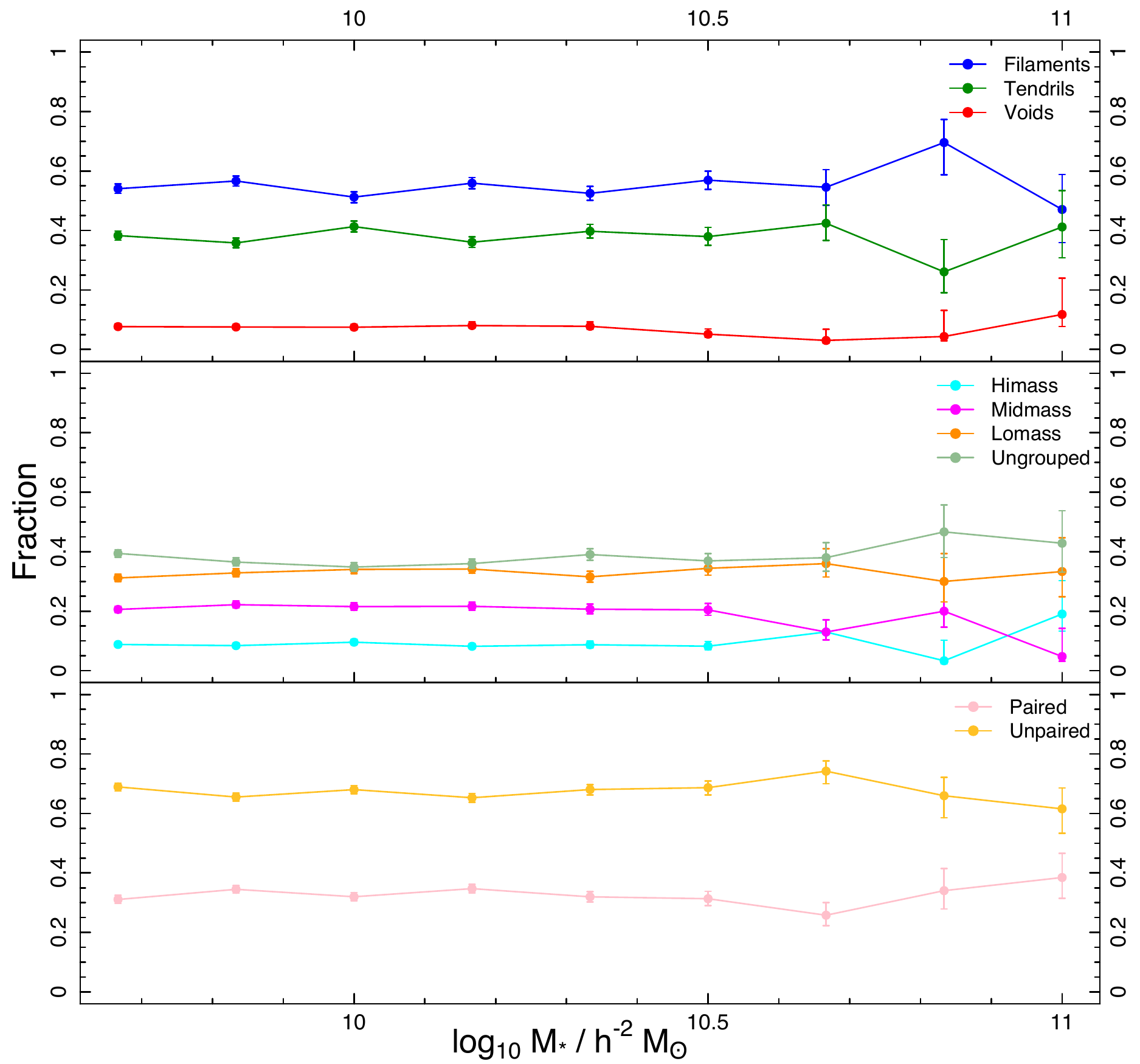}
	\caption{Fractional distribution of mass-matched galaxies in different environments as a function of stellar mass bins. Note that the fractions are calculated within each environment type (for example, the fraction of galaxies in pairs is given as $N_{\mathrm{pair}} / (N_{\mathrm{pair}} + N_{\mathrm{notpair}})$). Errors shown are calculated by sampling the Beta distribution \citep{Cameron2013}.}
	\label{fig:massfracbins}
\end{figure}

This mass segregation can be accounted for by resampling each environmental population of galaxies such that their galaxy stellar mass functions all match. In other words, we resample each population such that within the interval $[\widetilde{M}_*,\, \widetilde{M}_* + \ud \widetilde{M}_*$] is equal across all galaxy populations. We split each population into bins of size 0.05 dex of stellar mass between $9.5 \leq \log{M_*/h^{-2} M_{\odot}} \leq 11$, and in each bin, randomly resample the galaxies in each population (except for the voids) so that their GSMF in that bin matches that of the void population in that same bin. Galaxies whose mass is greater than $10^{11} \log{M_*/h^{-2} M_{\odot}}$ are discarded. The results of this process can be seen in Figure \ref{fig:massnorm}, where we plot the PDFs of stellar masses of galaxies in different environments. We choose to match all populations to the GSMF of the void galaxy population, as this is the most extreme (and because we cannot `upscale' that GSMF to the others). The stellar mass normalised samples are therefore random subsets of each galaxy sample that have the same mass distribution as the void sample. From here on, whenever we refer to a particular population of galaxies, we always refer to a sub-sample with mass matching; meaning that we are discussing galaxies whose stellar mass is between $9.5 \leq \log{M_*/h^{-2} M_{\odot}} \leq 11$. Figure \ref{fig:massfracbins} shows the relative fraction of galaxies in their environments, split by large scale structure, group membership, and pair membership.

For the remainder of this paper, we will combine our mass normalised environmental populations of galaxies with the various catalogues introduced in the preceding section. The sample we use is the galaxies that are common to all catalogues and are within our specified mass and redshift ranges. When mass normalised, this sample contains 28,781 galaxies. The sample size increases to 35,850 galaxies when the mass normalisation is removed. Note that all environmental categories of galaxies are mass matched independently.

\subsection{Colour, brightness, and morphology}

\begin{figure*}
	\centering
	\includegraphics[width=0.9\textwidth]{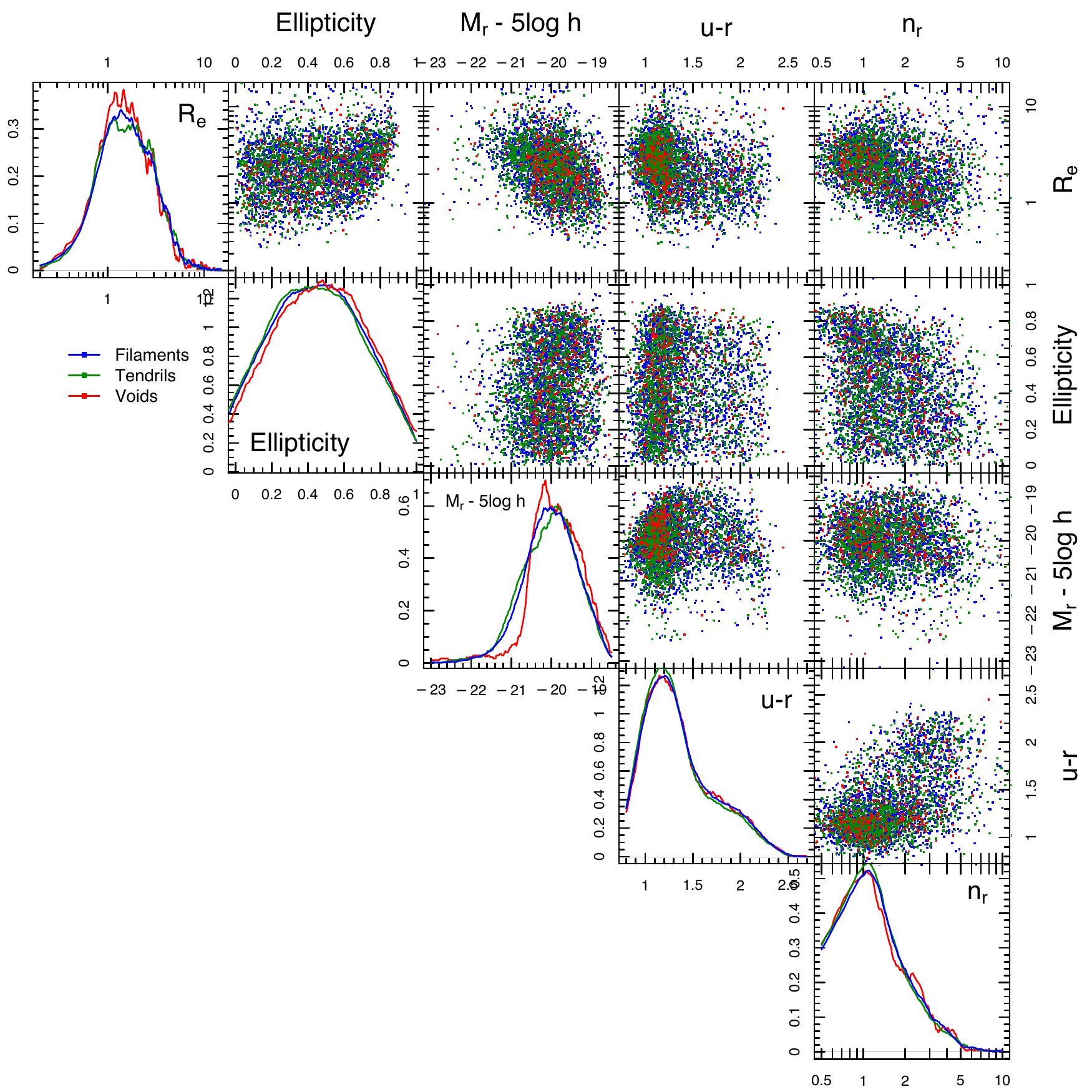}
	\caption{1 and 2D distributions of galaxies as a function of effective radius, ellipticity, absolute magnitude $M_r$, $u-r$ colour, and $r$-band S\'{e}rsic index $n_r$ for mass normalised galaxies in filaments, tendrils and voids (shown as blue, green and red lines and points respectively). Each frame plots two of these parameters against each other, while the histograms show the distributions of each one individually. Aside from very subtle differences in the distribution of ellipticity, effective radius, and absolute magnitude, there is very little difference in the properties of galaxies in the 3 environments.}
	\label{fig:bigmosaic}
\end{figure*}

Recent work studying the properties of void galaxies has shown them to be predominantly bluer and fainter with respect to their counterparts in higher density regions (e.g. \citealp{Kreckel2012}). Conversely, work by \citet{Eardley2014} which uses a tidal tensor description to classify GAMA galaxies into knots, filaments, sheets, and voids finds that the luminosity function of galaxies is independent from their large scale structure classification. In this work, we present a comparison of a number of visual and photometric properties of galaxies in Figure \ref{fig:bigmosaic}, where the $r$-band effective radius $R_e$, $r$-band ellipticity, $r$-band absolute magnitude $M_r$, rest-frame and extinction corrected $u-r$ colour, and $r$-band S\'{e}rsic index $n_r$ of galaxies in filaments, tendrils, and voids (in blue, green, and red respectively) are compared to each other. The figure is arranged as a grid, with the scatterplots showing relationships between these various parameters, while the 1D histograms at the end of each row represent the distribution of that single parameter for the three populations of galaxies.  

When we produce our mass normalised samples, we do not attempt to normalise for location within groups (i.e., dark matter halos), so it is plausible that the fraction of satellite galaxies differs between the various samples. However, as discussed below, this does not appear to produce artificial differences between the void, tendril and filament samples.

No population of galaxies uniquely inhabits a particular corner of the parameter spaces we are studying (for example, filament galaxies are not exclusively red, and do show some signal of bimodality in colour). In the context of the properties shown in this figure, large scale structure has little impact on galaxy properties, beyond determining the form of the GSMF which is an important environmental effect in itself. There are no statistically significant differences between the colour and S\'{e}rsic index of galaxies in filaments, tendrils, and voids; performing a two-sample KS-test between the $u-r$ and $n_r$ values between all three populations yields P values far exceeding 0.05, indicating that  they are consistent with having been drawn from the same distribution. We only see subtle differences with respect to their effective radius and ellipticity. The difference in ellipticity might therefore be an indicator of discs being more prevalent in voids. The absolute magnitudes of void galaxies suggests that they are slightly, but consistently fainter than galaxies in filaments and tendrils. Given that this is a mass controlled sample, this implies a systematically higher mass-to-light ratio for void galaxies and is consistent with the picture of galaxies in more dense environments being able to acccumulate a larger, brighter stellar population.  

\subsection{Total energy output}
\label{sec:csed}

The cosmic spectral energy density (CSED, \citealp{Driver2013}) describes the total energy being generated by galaxies within a representative volume of the Universe at a given epoch. The CSED provides an empirical measurement of the energy budget of a population of galaxies by wavelength, which is instrumental in understanding the different physics and stellar populations of different types of galaxies. An attractive feature of the CSED is that it can be modeled (e.g. \citealp{Somerville2012,Driver2013a}) given \emph{a priori} knowledge of the cosmic star formation history, initial mass function, a stellar population model, and a dust attenuation model. Following the prescription of \citet{Driver2013}, we construct CSEDs of galaxies in filaments, tendrils, and voids across 20 bands in the FUV-FIR range. We focus on this wavelength range as it is entirely dominated either by direct starlight (FUV to NIR) or reprocessed starlight that has been re-emitted by dust (FIR). At $z \leq 0.1$ the impact of AGN at these wavelengths is neglible \citep{Driver2008,Driver2013}.

\begin{figure}
	\centering
	\includegraphics[width=0.5\textwidth]{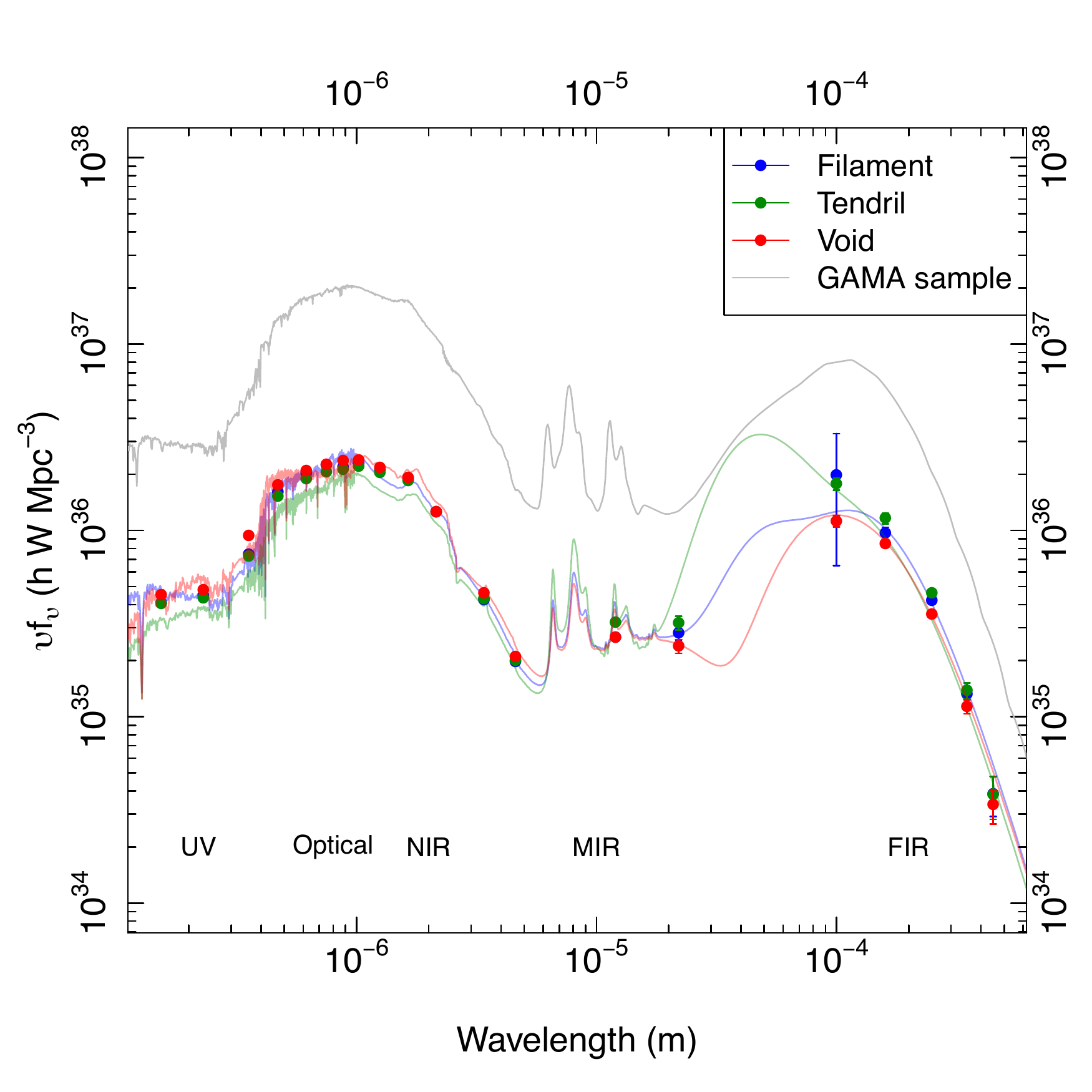}
	\caption{Total flux densities in 20 bands for mass normalised galaxies in filaments, tendrils, and voids, shown in blue, green and red respectively. The CSED shown in grey has been taken from \citet{Driver2013} and shows an example of a typical GAMA galaxy SED. It has been scaled to lie above the points.}
	\label{fig:CSED}
\end{figure}

We construct our CSEDs by first summing up the luminosities of galaxies with photometry in all 20 bands that reside in filaments, tendrils, and voids; we consider only galaxies that are observed in all bands. This sum gives the luminosity density of those galaxies, which can then be converted to an observed energy density $\epsilon_{\mathrm{obs}}$ by multiplying by the effective mean frequency of each filter, or band. The intrinsic energy density is then $\epsilon_{\mathrm{int}} = \epsilon_{\mathrm{obs}} / p_{\mathrm{esc},\lambda}$, where $p_{\mathrm{esc},\lambda}$ is the mean photon escape fraction for each band, derived as in \citet{Driver2007} using the dust attenuation model of \citet{Tuffs2004} (see also \citealp{Popescu2011}). Finally, we normalise the energy densities in all bands to that of the $K$-band, and show the results in Figure \ref{fig:CSED}. The data points represent the flux in each band, and we use MAGPHYS (Multi-wavelength Analysis of Galaxy Physical Properties; \citealp{DaCunha2012}) to fit an unattenuated CSED to each data set; these fits are designed to be for illustrative purposes only.

Figure \ref{fig:CSED} shows some differences between the energy outputs of galaxies in different large scale environments. Void galaxies show a very subtle decrease in emission in the FIR/submm bands. It is clear from the fits shown in this figure that the dust content is very poorly constrained, and so we cannot make statistically significant statements about the meaning of this small difference.

\subsection{Metallicity}

Measuring the abundance of metals in the gas of galaxies is a strong indicator of the chemical reprocessing that has taken place in their stars, as well as the presence of heavier elements in their interstellar medium that have accumulated as a result of previous generations of star formation. Different elemental abundances can be affected by a number of dynamical processes, and are sensitive to environment, as well as stellar mass (e.g. \citealp{Foster2012,Lara-Lopez2013}). Recent results show that galaxies in clusters have a tendency to be slightly more metal-rich \citep{Cooper2008}, but that this relationship is largely driven by galaxies in very high density regions and local environment as opposed to cluster membership \citep{Ellison2009}. On the other hand, \citet{Hughes2013} find that in a sample of 260 nearby late-type galaxies, the stellar mass-metallicity relation is insensitive to environment. 

\begin{figure}
	\centering
	\includegraphics[width=0.5\textwidth]{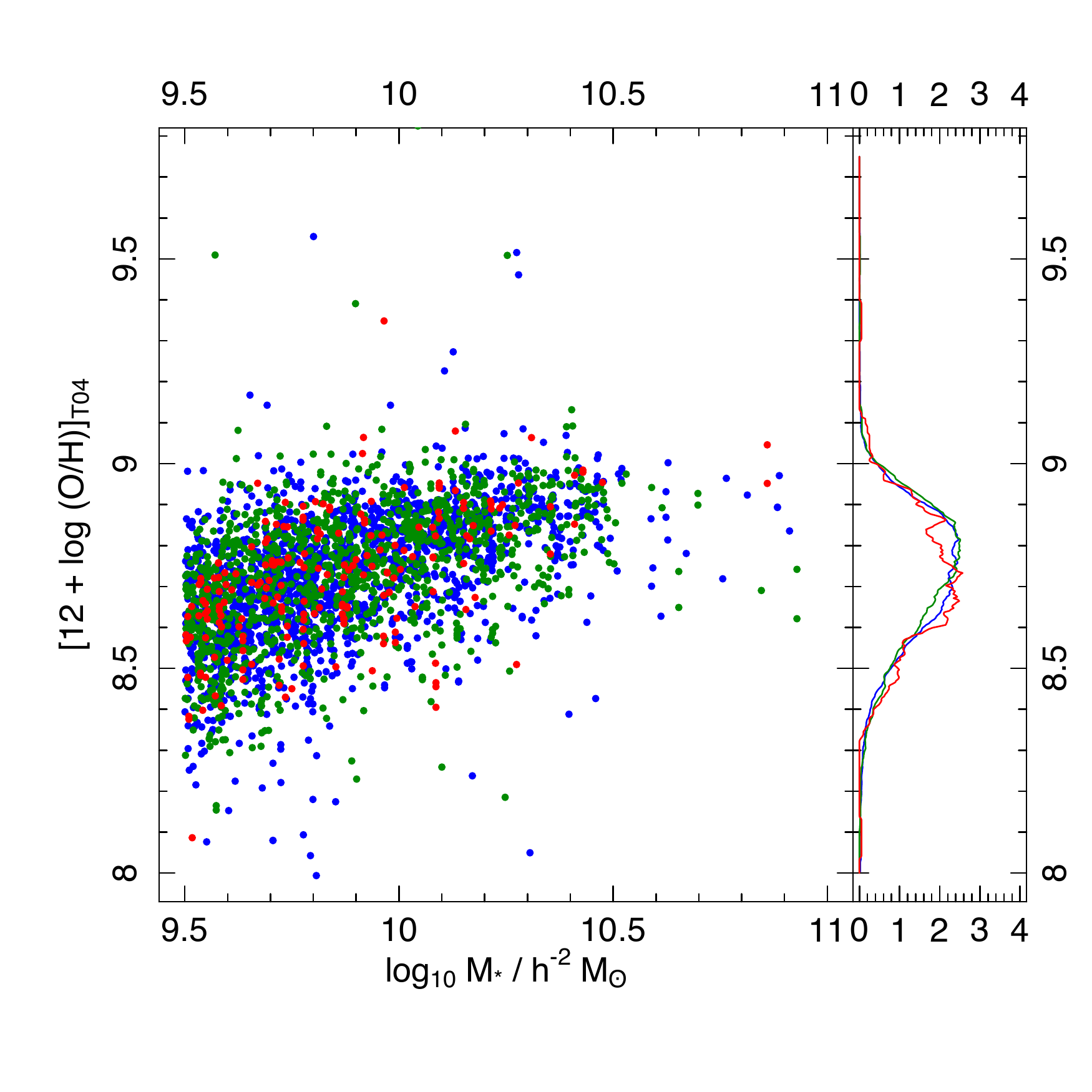}
	\caption{Metallicities and stellar masses from \texttt{EmLinesGAMAIIv01} and \texttt{StellarMassesv16} shown for mass normalised filament, tendril and void galaxies in blue, green and red respectively.}
	\label{fig:metalcat}
\end{figure}

We examine the mass-metallicity relation for star forming galaxies in filaments, tendrils, and voids using gas metallicities. While there is no notable difference between filament and tendril galaxies, we do find that void galaxies have subtly lower metallicities in comparison to these other two populations, as can be see on the right hand panel of Figure \ref{fig:metalcat}. We suggest therefore that material in filaments and tendrils may be being processed in a similar way, and perhaps less efficiently in void galaxies. However, considering the similarities in their CSEDs as shown in Figure \ref{fig:CSED}, this effect is not thought to be significant.

\subsection{Comparison to group and pair classifications}

It is instructive to repeat some of the above analysis on galaxies classified according to their presence within a group, or a galaxy-galaxy pair. This allows us to gain some understanding into a `hierarchy' of how environment affects galaxy evolution: from pair-pair interactions, to the mass of the halo or group in which the galaxy resides, to its presence within large scale structure.

We take our original sample of galaxies and classify them not by large scale structure, but by presence within a group. If the galaxy is in a group, then it is classified according to the mass of the group (taken from \texttt{GroupFindingv07}). These are dynamical mass estimates of each group calculated using the group's velocity dispersion and radius. See \citet{Robotham2011} for a discussion on how these properties were measured. We also identify all galaxies in pairs that are within 100 h$^{-1}$ kpc from each other, with velocity separations of $\leq 1000$ kms$^{-1}$ \citep{Robotham2014}. 50.16\% of galaxy pairs in groups exist within groups with 3 or more members; in other words, they are pairs within larger groups. Finally, note that all samples are mass matched to the void galaxy population.  

\begin{figure*}
	\centering
	\includegraphics[width=0.9\textwidth]{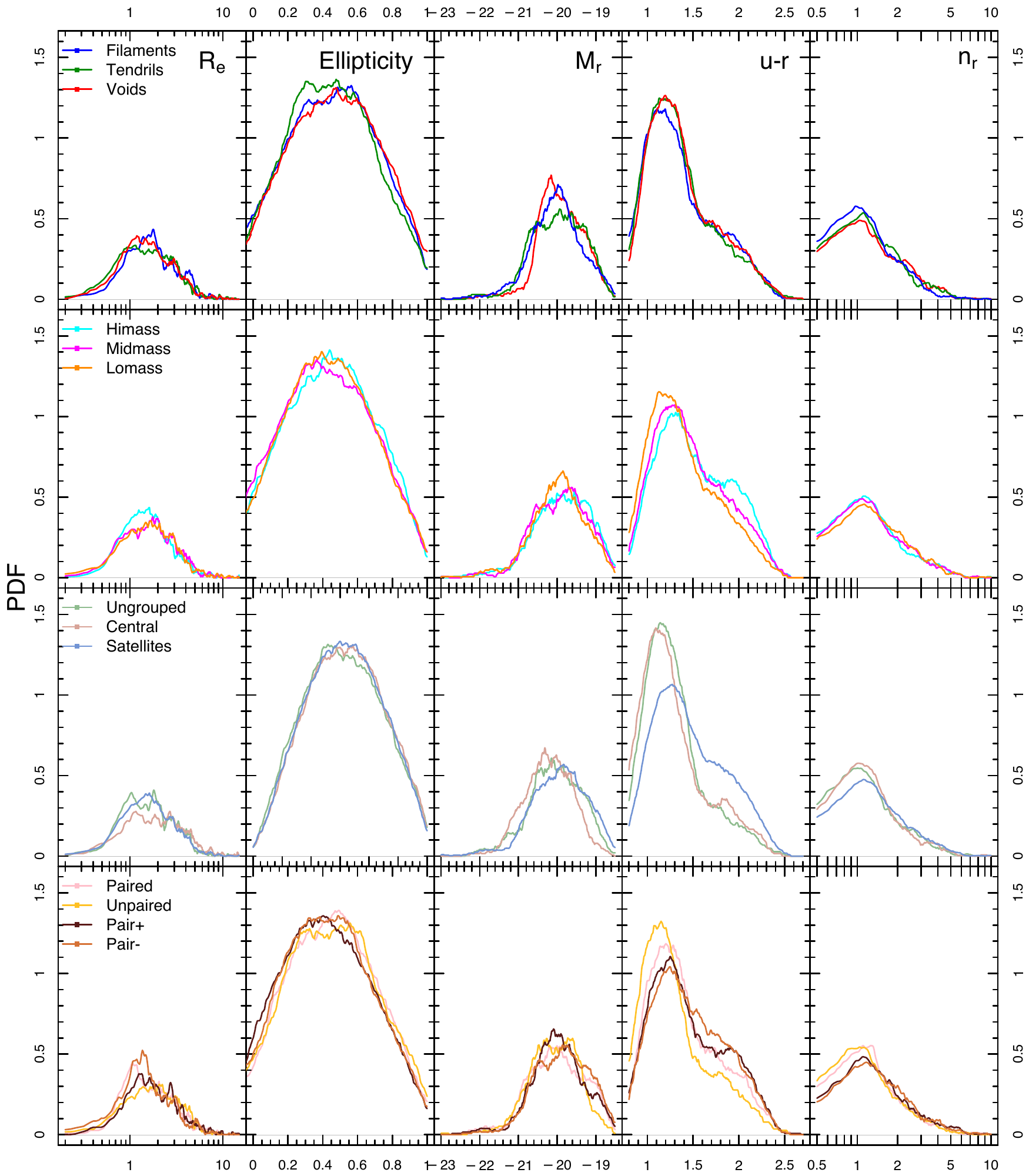}
	\caption{The probability distribution functions of effective radius, ellipticity, $r$-band absolute magnitude $M_r$, $u-r$ colour, and S\'{e}rsic index $n_r$ for galaxies split by three different environment types: large scale structure (top row), halo mass and galaxies not in groups (second row), and paired and unpaired galaxies (bottom row). There is an overall trend for galaxies outside of groups, in group centres, and outside of pairs to show similar characteristics, particulary in $R_e$, $u-r$ colour, and $n_r$.}
	\label{fig:comphist}
\end{figure*}

We take the same properties used to generate Figure \ref{fig:bigmosaic} and this time plot them for galaxies in groups in high mass, mid mass, and low mass halos (defined as $M_H \geq 10^{14} h^{-1} M_{\odot}$, $10^{13} \leq M_H \leq 10^{14} h^{-1} M_{\odot}$ and $M_H \leq 10^{13} h^{-1} M_{\odot}$ respectively), galaxies not in groups, central and satellite galaxies in groups, galaxies within and outside pairs, and the dominant and sub-dominant galaxy of each pair (all of these populations are mass controlled, as described in Section \ref{sec:massnorm}). Figure \ref{fig:comphist} displays the distribution of these parameters as a function of different environments, with each row representing a different classification type. From top to bottom, the rows show classifications by large scale structure, presence in a group, and presence in a pair. Pair+ and Pair- denote the dominant and sub-dominant member of a pair.

Figure \ref{fig:comphist} reveals a number of interesting trends: namely that different environments tend to separate out certain galaxy properties better than others. An obvious example of this is that halo mass (or group membership) does not appear to be a strong predictor of galaxy luminosity, or ellipticity. Similarly, halo mass alone does not strongly correlate with S\'{e}rsic index. There is an overall tendency for galaxies that are not in pairs, and not in groups, to exhibit similar properties: fainter, bluer, higher ellipticity, and a S\'{e}rsic index consistent with being a disc galaxy. On the other hand, galaxies that reside within groups and pairs have a bimodal colour distribution with a higher red peak, higher S\'{e}rsic indices consistent with early type galaxies, but do not tend to be more luminous; though we make this statement whilst acknowledging the caveat that we have controlled our sample for mass. 

Our results regarding the properties of pairs are in excellent agreement with \citet{Robotham2013a}, who conclude that the net effect of close pairs is a supression of star formation. This is reflected in our results as paired galaxies being far redder than unpaired galaxies (which is also in good agreement with Davies et al., in prep). These results suggest a scenario where the process of infalling into a higher density structure generates changes in the morphology and colour of a galaxy. This is consistent with the recent results (e.g. \citealp{Valentinuzzi2011,Lopes2013,Robotham2013a,Lacerna2014}) which indicate that local density plays a more important role in galaxy evolution than the mass of the halo in which the galaxy resides. Central galaxies in groups occupy the same parameter space as ungrouped galaxies; this is consistent with the findings of \citet{Robotham2013a}. Ungrouped galaxies can be interpreted as the central galaxies of undetected groups (i.e. GAMA is not sufficiently deep to observe their satellites). Notably, Figure 10 of \citet{Robotham2013a} establishes that central galaxies with $M_* \approx 1.25 \times 10^{10} h^{-2}$ M$_{\odot}$ are more likely to be late type than centrals of a higher mass. The distribution of S\'{e}rsic indices for central galaxies in our sample of $9.5 \leq \log{M_*/h^{-2} M_{\odot}} \leq 11$ is consistent with this finding, and suggests that central galaxies with even lower masses are even more likely to be late type. It is the satellite galaxies in groups that are most directly affected by the group environment, as shown in their distribution in Figure \ref{fig:comphist}. Again, this is consistent with the findings of \citet{Robotham2013a}.

\begin{figure*}
	\centering
	\includegraphics[width=1.0\textwidth]{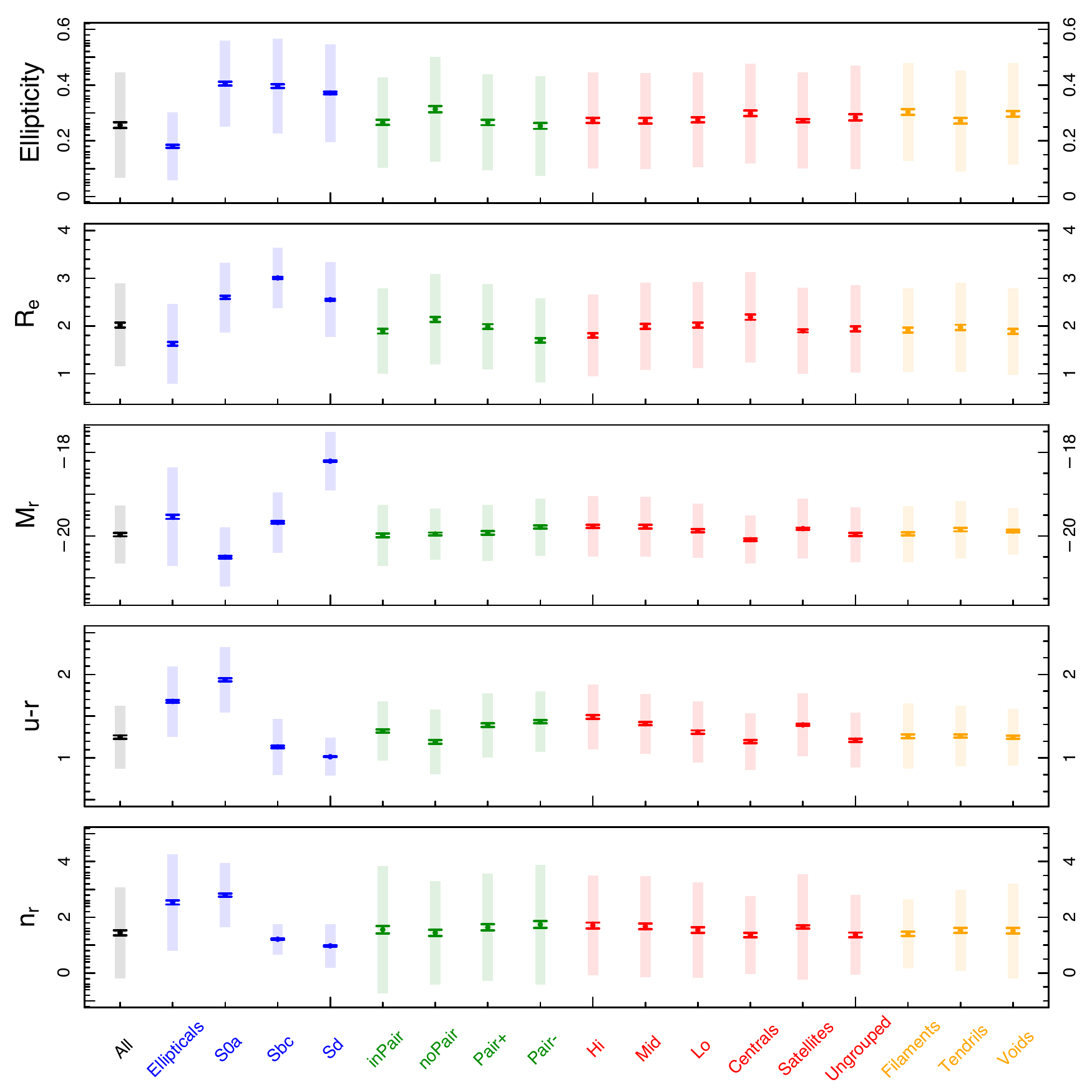}
	\caption{Distributions of the same properties shown in Figure \ref{fig:bigmosaic}, but for a greater number of populations. The points plotted are medians of that parameter for each population, with shaded regions showing the standard deviation of each parameter, and the error bars representing the standard error of the mean (i.e. SD/$sqrt(N-1)$. It is for galaxies classified by morphological type that we see the greatest variation in properties; all other populations span the same approximate range (as shown by the vertical lines) but do show differences in their medians and the widths of their distributions.}
	\label{fig:sidewaysPlot}
\end{figure*}

Figure \ref{fig:sidewaysPlot} shows an alternative view of the properties shown in previous sections, and for a wider selection of environments. Alongside the environment types discussed above, for this analysis we also consider visual classification of morphology, i.e. E, S0a, Sbc, and Sd galaxies. These galaxies were visually classified by three pairs of independent observers as being early type or late type, based on the presence of a disc or disc-like feature. The sample contains approximately 20,000 galaxies, from $0.002 < z < 0.1$. All environment and morphological samples are mass matched to the void galaxies. The environment types are colour coded according to common `types': morphological types are shown in blue; pairs in green; groups in red; and large scale structure in orange. Shaded regions show the spread of the data, and the error bars are the standard error about the median.

\subsubsection{Elliptical fraction}

\begin{figure}
	\centering
	\includegraphics[width=0.5\textwidth]{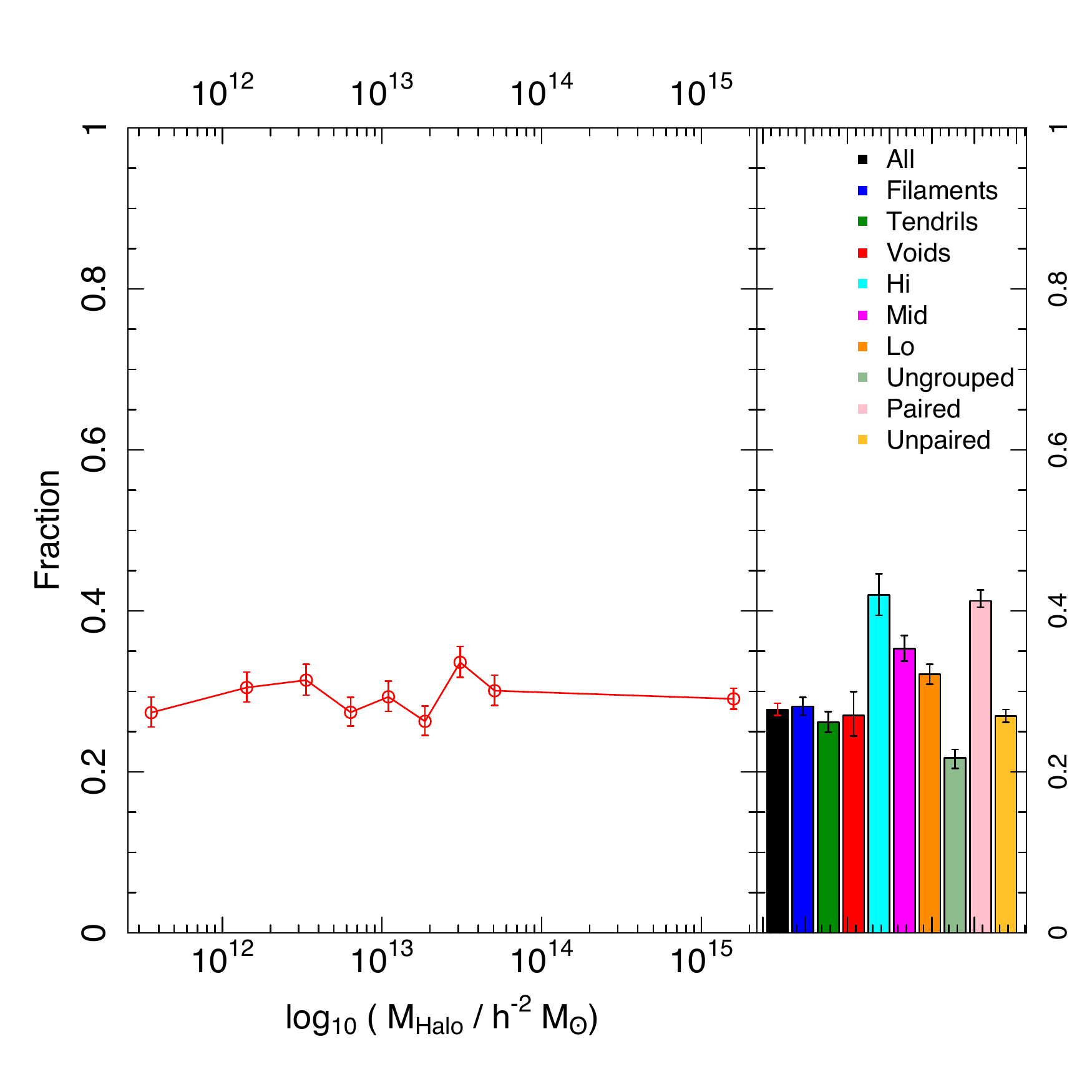}
	\caption{Fraction of ellipticals of \emph{all} galaxies in \texttt{VisualMorphologyv02} that are in groups as a function of halo mass shown in red. The fraction of ellipticals does not change as a function of group mass. Note that elliptical fractions are calculated in bins of halo mass each containing an equal number of galaxies. The right hand panel shows the fraction of elliptical galaxies in a variety of environments. The error bars show $1\sigma$ uncertainties in the population fraction, as calculated using the \textsc{qbeta} function \citep{Cameron2013}.}
	\label{fig:ellipfrac}
\end{figure}

We can similarly look at the fractions of early and late type galaxies in these environments, using the visually classified morphological galaxy sample. We identify all galaxies in this sample that belong to a group, and track the fraction of early type galaxies in groups as a function of halo mass. These results can be seen in Figure \ref{fig:ellipfrac}, where the left panel shows, in red, the fraction of early type galaxies as a function of halo mass. We do not see a correlation between early type fraction and group mass, when considering all galaxies (i.e. without mass normalisation).

\begin{figure}
	\centering
	\includegraphics[width=0.5\textwidth]{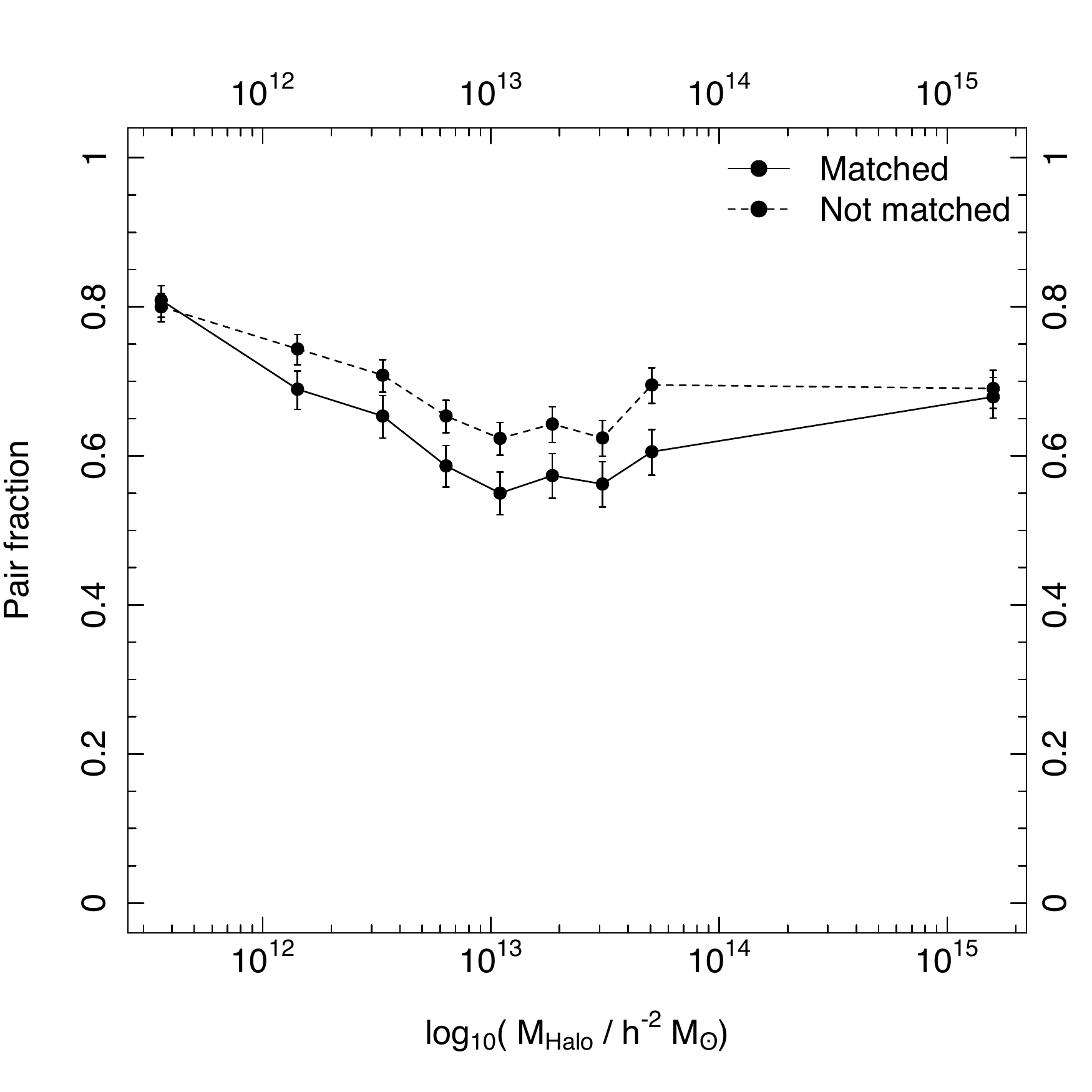}
	\caption{Fraction of galaxies that are considered to be in pairs as a function of halo mass, in bins of equal size. For each bin we plot the fraction $N_{\mathrm{pair}} / (N_{\mathrm{pair}} + N_{\mathrm{notpair}})$ of all galaxies in halos of that mass bin, and include both mass matched and unmatched galaxies. We note a slight tendency for the fraction of galaxies in pairs to fall in higher mass halos; and that this trend persists even when we remove the mass matching.}
	\label{fig:pairfrac}
\end{figure}

The right hand panel of Figure \ref{fig:ellipfrac} displays the total fraction of early type galaxies in different environment types. These results are consistent with the galaxy types identified in Figure \ref{fig:comphist}, in the sense that galaxies in the field have a lower early type fraction. Unlike the left panel of the figure, which shows the elliptical fraction as a function of halo mass for all galaxies, the right hand panel shows that mass-controlled galaxies in high mass halos have a higher elliptical fraction than those in mid and low mass halos. Given that the upward trend in the fraction of early type galaxies with halo mass is seen only for a narrow subset of mass controlled galaxies, we reason that the presence of massive ellipticals in halos of all masses drives up the fraction in the left hand panel. To check if the increase in elliptical fraction in high mass groups is simply due to a larger presence of pair galaxies in such systems, we show the fraction of galaxies in pairs in our mass normalised sample as a function of halo mass in Figure \ref{fig:pairfrac}. We note a small, statistically significant downward trend in pair fraction as a function of halo mass. Figure \ref{fig:pairfrac} indicates that the increased fraction of elliptical galaxies in more massive halos (for a mass controlled sample) is not related to presence in pairs.

\section{How does mass normalisation affect our results?}

Our analysis and results so far have been restricted to the narrow and noisy mass distribution occupied by our sample of void galaxies. In this section, we not only examine the properties of galaxies without any mass control, but also match them to the smoother \citet{Baldry2012} GSMF.

\subsection{Removing mass control}

\begin{figure*}
	\centering
	\includegraphics[width=1.0\textwidth]{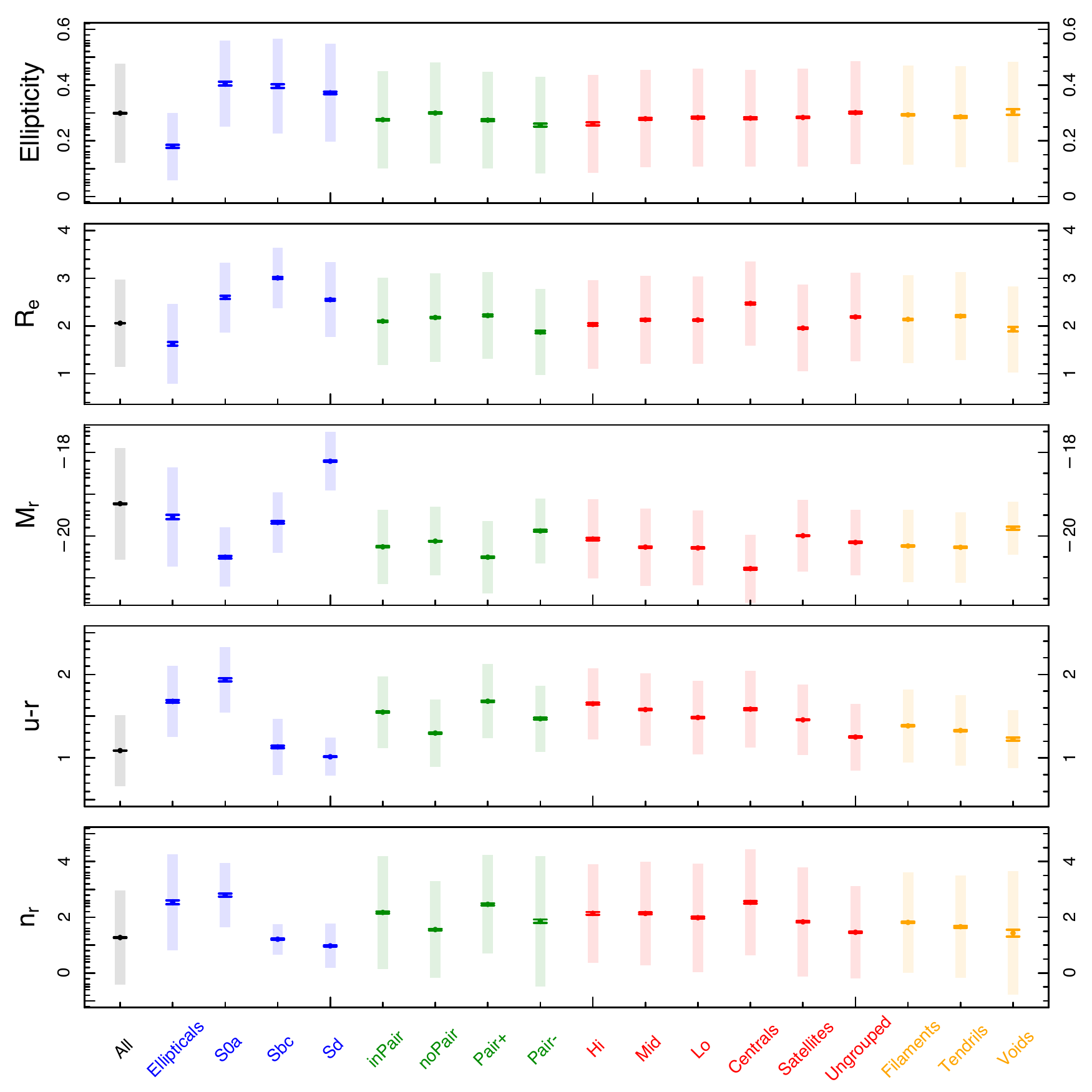}
	\caption{As with Figure \ref{fig:sidewaysPlot}, but for galaxies whose masses have not been normalised. Most trends are notably more exaggerated in this figure, with the exception of the morphologically classified galaxy population, which is a strong indicator for the dominance of stellar mass and local density in influencing these parameters.}
	\label{fig:sidewaysPlot_notNorm}
\end{figure*}

In order to better understand the effect of controlling stellar mass in our sample, we replicate the results of Figure \ref{fig:sidewaysPlot} for galaxies without mass normalisation. Removing mass normalisation has a number of effects, which can be seen in Figure \ref{fig:sidewaysPlot_notNorm}. An immediately obvious result is that the galaxy properties shown in this figure show greater trends with environment (with the exception of those for galaxies classified by morphological type), where we continue to see the same trends. If the reintroduction of stellar mass variations leads to greater trends in the properties of galaxies classified by environment and \emph{not} by morphology, it stands to reason that stellar mass is more influential in governing the parameter space that galaxies inhabit, cf. large scale environment. A secondary conclusion drawn from Figure \ref{fig:sidewaysPlot_notNorm} is that morphological classifications of galaxies are strong predictors of their properties, even when controlling for stellar mass.

\begin{figure*}
	\centering
	\includegraphics[width=1.0\textwidth]{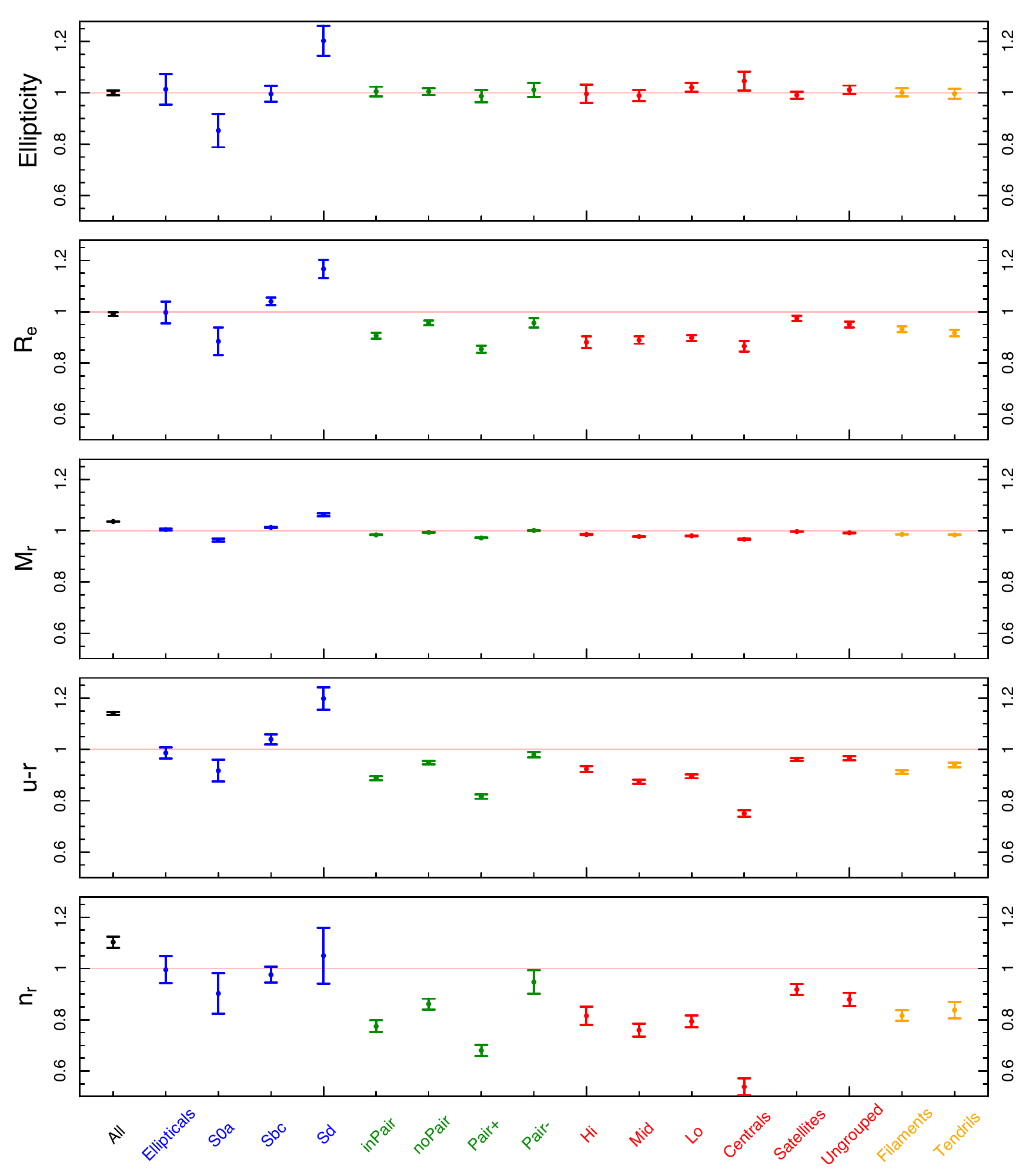}
	\caption{Ratios of the median values of each of two galaxy samples shown in Figures \ref{fig:sidewaysPlot} and \ref{fig:sidewaysPlot_notNorm} (i.e. the ratio between the median for the mass normalised sample and the non-mass normalisation sample). The errors shown are standard errors about the median. We do not show void galaxies in this Figure, as the population is the same for both samples.}
	\label{fig:sidewaysPlot_diff}
\end{figure*}

We can also compare the medians of the distributions of each property shown in Figure \ref{fig:sidewaysPlot_notNorm} between the mass normalised and non-mass normalised populations. Such a comparison quantitatively highlights the effects of mass normalisation. In Figure \ref{fig:sidewaysPlot_diff} we plot the ratios of the medians of the mass normalised and non-mass normalised galaxies for each population, for each of the properties discussed in previous figures. The void galaxy sample is the same for both samples, and so is not shown in this Figure. Different populations appear to show a different amount of susceptibility to mass normalisation, as evidenced by the larger value of the ratio in those properties. Colour and S\'{e}rsic index appear to be the most affected. Central galaxies exhibit the largest departure from unity, once again highlighting how the properties of central group galaxies differ with mass, as shown in \citet{Robotham2013a}. 

\begin{figure*}
	\centering
	\includegraphics[width=1.0\textwidth]{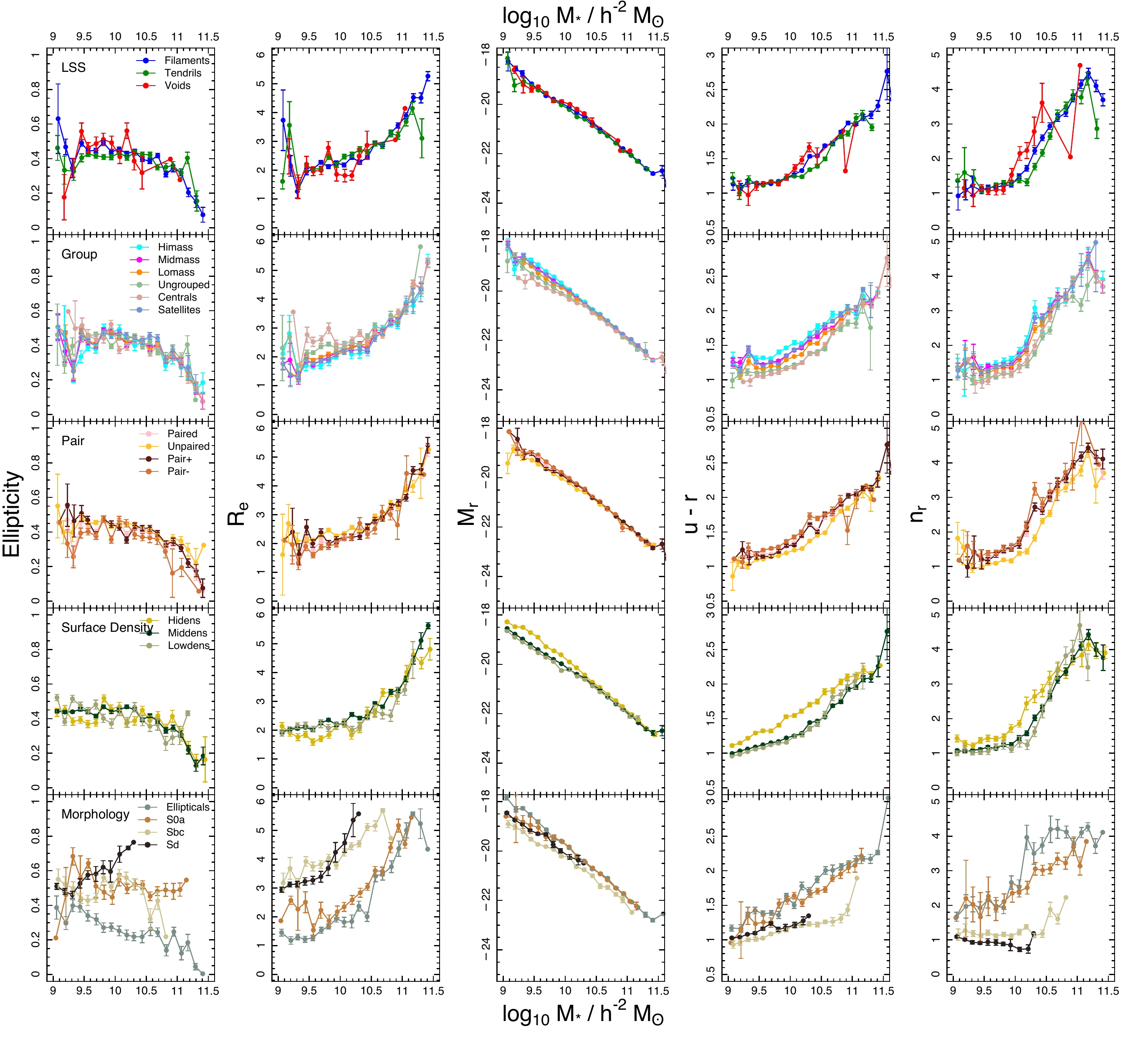}
	\caption{Relationships between stellar mass and ellipticity, r-band effective radius $r_e$, absolute magnitude $M_r$, $u - r$ colour, and r-band S\'{e}rsic index $n_r$ for galaxies classified according to large scale structure, group membership, pair membership, surface density, and morphology. Each population is binned in stellar mass, and the x-coordinate of each bin is the average stellar mass of the galaxies in that bin. Errors shown are standard errors about the mean (points with no error bars are single detections). There is an overall tendency for galaxies in all environments to be corellated with mass in a similar way, with only morphological classifications showing distinct trends.}
	\label{fig:bigBinGrid}
\end{figure*}

To further explore these results, we study the mass evolution of the parameters examined in Figure \ref{fig:sidewaysPlot_notNorm} across environments. We include all galaxies that were used in generating the GSMFs in Figure \ref{fig:stellarMassFunc}, without any mass matching in any populations. The relationship between stellar mass and these parameters (ellipticity, r-band effective radius $r_e$, absolute magnitude $M_r$, $u - r$ colour, and r-band S\'{e}rsic index $n_r$) is shown in Figure \ref{fig:bigBinGrid}. We also include this relationship for galaxies classified by their local density (based on the distance to the 5th nearest neighbour of each galaxy). We split this population into 3 bins: containing the lowest $1\sigma$ of the data, between $1\sigma$ and $3\sigma$, and above $3\sigma$. Galaxies in large scale structure, groups, and pairs tend to to show a similar evolution in their parameters with stellar mass regardless of the environment they reside in. We do see a tendency for galaxies in more massive halos, and in pairs, to be systematically more red and have a higher S\'{e}rsic index across the whole mass range, but this pattern is not seen for large scale structure. Galaxies classified by morphology display considerably different trends. We should not be surprised that this is the case for ellipticity, $n_r$, and $r_e$ (although it is notable that $n_r$ tends to increase with mass for galaxies classified as ellipticals or S0a galaxies). 

\subsection{Baldry 2012 GSMF}

\begin{figure}
	\centering
	\includegraphics[width=0.5\textwidth]{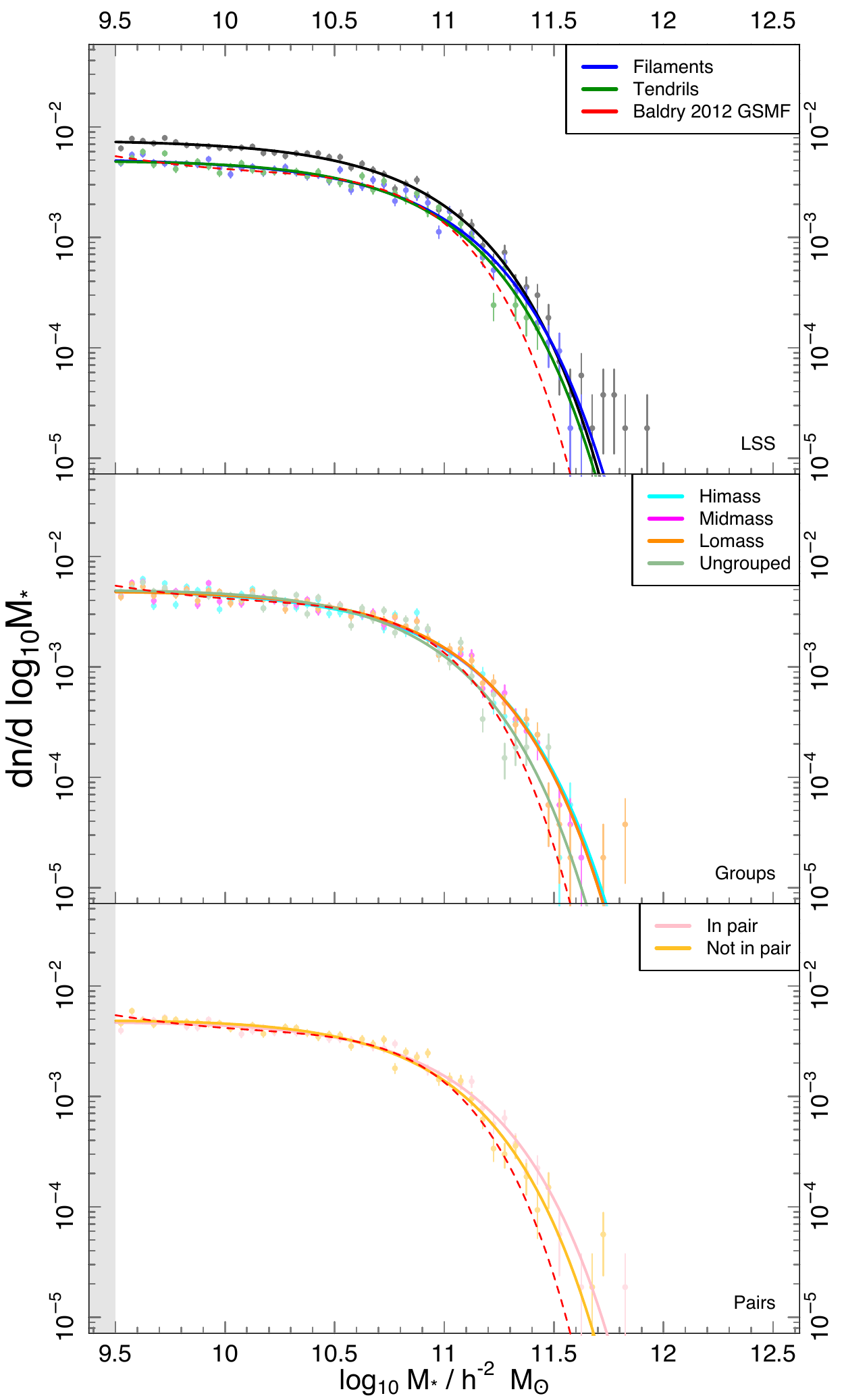}
	\caption{GSMFs of galaxies separated by environment, as in Figure \ref{fig:stellarMassFunc}; this time compared to the GSMF from \citet{Baldry2012}, which is shown as the dashed magenta line in each panel.}
	\label{fig:baldryGSMF}
\end{figure}

The \citet{Baldry2012} GSMF describes the mass distibution of 5210 galaxies from the first three years of the GAMA survey, covering an area of 143 square degrees. The galaxies are selected such that they are magnitude complete ($r = 19.4$ mag for galaxies in G09 and G15 and $r = 19.8$ mag for G12) and with $0.002 < z < 0.06$. The GSMF is then computed using a density corrected $1/V_{\mathrm{max}}$ method (see Section 2.5 of \citet{Baldry2012} for details). We show this GSMF as the dashed magenta line in each panel of Figure \ref{fig:baldryGSMF}, where the mass functions of galaxies in different environments, sampled such that they now match to this GSMF are also shown. Void galaxies are necessarily removed from this analysis. These stellar mass functions have slightly higher number densities at higher masses; this is caused by small numbers of high mass ($\geq 10^{11.5} h^{-2} M_\odot$) galaxies influencing our Schechter function fits. Across all environments, our sample contains only 17 such high mass galaxies, so this discrepancy has no discrenible influence on our conclusions.

\begin{figure*}
	\centering
	\includegraphics[width=1.0\textwidth]{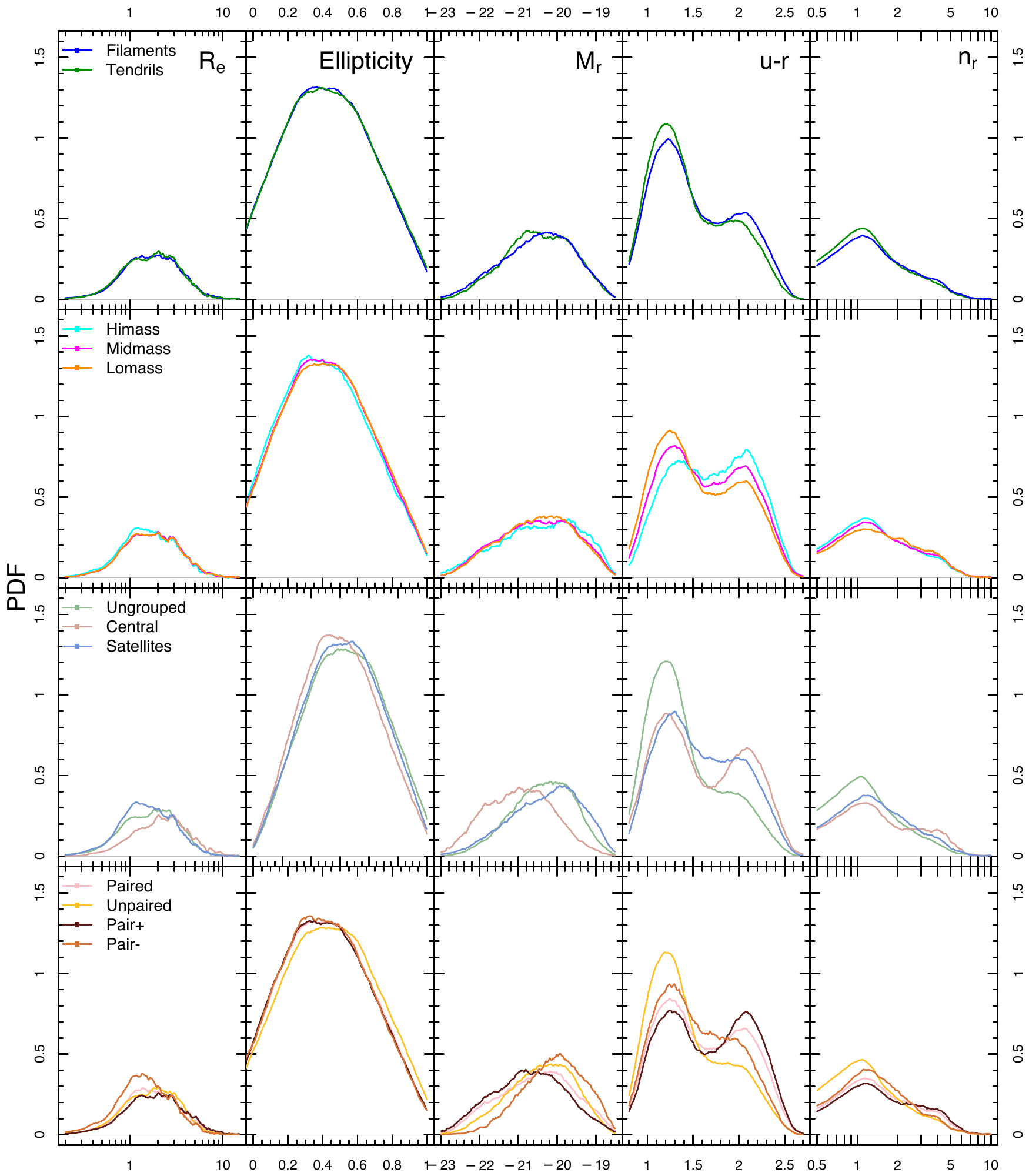}
	\caption{As with Figure \ref{fig:comphist}, but for galaxies matched to the \citet{Baldry2012} GSMF.}
	\label{fig:comphist_gsmf}
\end{figure*}

\begin{figure*}
	\centering
	\includegraphics[width=1.0\textwidth]{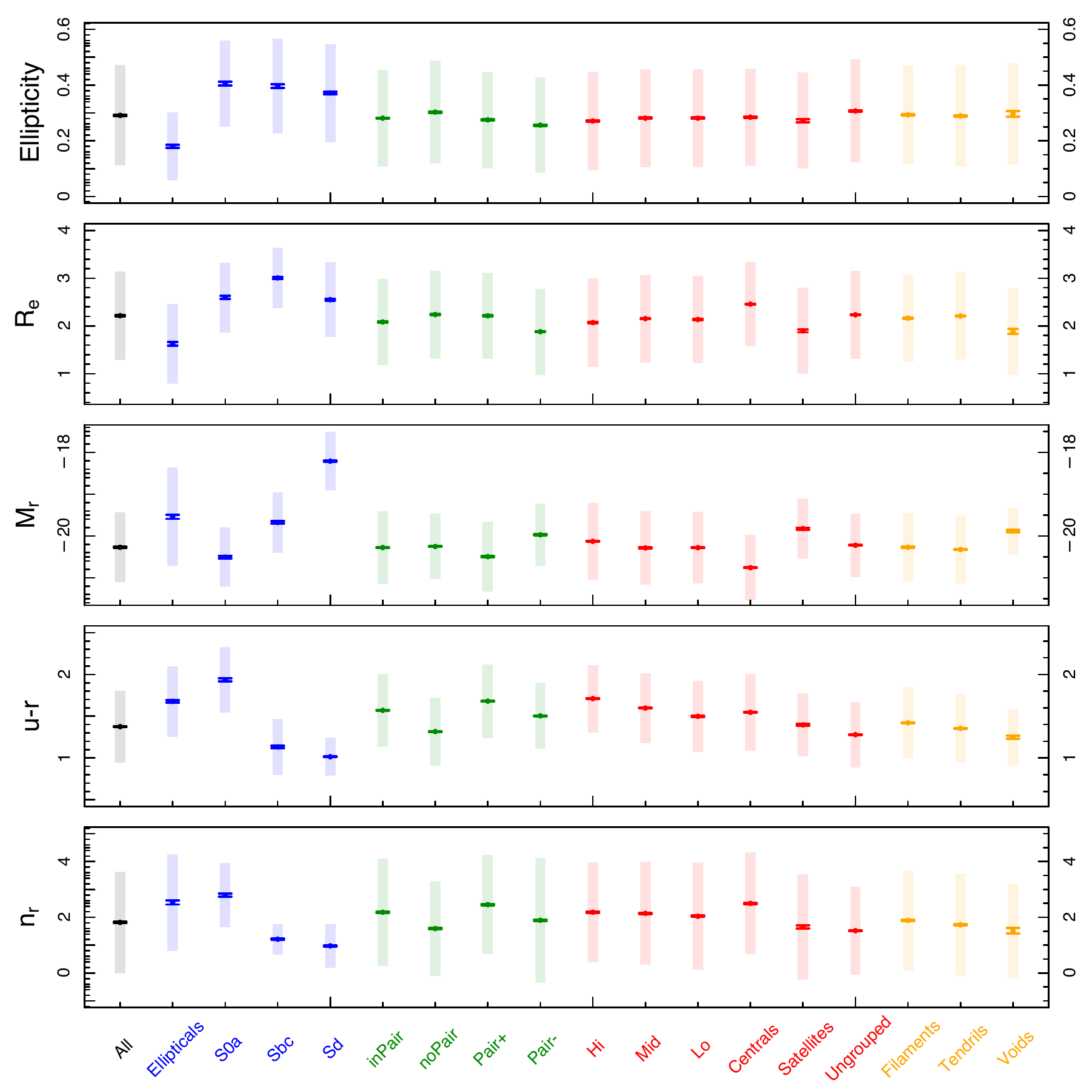}
	\caption{As with Figures \ref{fig:sidewaysPlot} and \ref{fig:sidewaysPlot_notNorm}, but for galaxies matched to the \citet{Baldry2012} GSMF.}
	\label{fig:sidewaysPlot_gsmf}
\end{figure*}

The resulting mass matched sample is similar to the void matched sample used in previous sections of this paper in that it provides a mass-unbiased selection of galaxies. By matching to the \citet{Baldry2012} GSMF; however, we are able to include a larger number of galaxies at higher mass; most importantly, galaxies whose mass is greater than $M_*$. It is therefore a high mass alternative to the void matched sample used earlier in this paper. A version of Figure \ref{fig:comphist} made with this higher mass matched sample is shown in Figure \ref{fig:comphist_gsmf}. Similarly, Figure \ref{fig:sidewaysPlot_gsmf} is equivalent to Figures \ref{fig:sidewaysPlot} and \ref{fig:sidewaysPlot_notNorm}. Large scale structure (i.e. filaments and tendrils) continue to have similar characteristics at these higher masses. For groups and pairs, the inclusion of higher mass galaxies has a number of discernible effects in the distribution of properties. While group mass continues to be a poor predictor of $M_r$ and $n_r$, centrals, satellites, and ungrouped galaxies show a greater variation in their luminosities compared to galaxies in groups. All populations show an increase in the bimodality of their colours, as a result of the inclusion of higher mass galaxies. Central galaxies are now less likely to have S\'{e}rsic indices consistent with having a disk, which is consistent with the findings of \citet{Robotham2013a}, who show that higher mass central galaxies are more likely to be elliptical. Dominant and sub-dominant pair galaxies show a large difference in their $M_r$ distributions compared to Figure \ref{fig:comphist}, and we see fewer blue dominant pair galaxies. These results are also consistent with \citet{Robotham2013a}, who show that more massive dominant pair galaxies are more likely to be early type than their lower mass dominant pair counterparts.

\section{Summary and conclusion}

We have used structural and photometric data products from the GAMA survey to seek out trends in galaxy properties as a function of nonlocal environment. We focus in particular on parameters directly affected by changes in stellar populations: colour, brightness, morphology, and total energy output, as well as some derived parameters. Figure \ref{fig:stellarMassFunc} highlights the fact that galaxies in different environments have strongly varying stellar mass functions. We control for this aspect by normalising our samples so that their stellar mass functions match those of void galaxies, which have the most extreme mass distribution. This ensures that our comparisons are free from any stellar mass bias. By looking at the total CSED of galaxies in different environments in Figure \ref{fig:CSED} we can see that void galaxies have slightly lower emission in IR bands, but these fits are poorly constrained. Similarly, metallicity does not appear to be strongly affected by large scale structure (Figure \ref{fig:metalcat}). Figure \ref{fig:bigmosaic} establishes that when controlling for the mass distribution, the properties of galaxies in different large scale environments are largely the same.

We broaden our investigation and look at how the properties examined in Figure \ref{fig:bigmosaic} change for galaxies in group and pair environments. Our results are shown in Figure \ref{fig:comphist} and \ref{fig:sidewaysPlot}, and suggest a tendency for galaxies outside of structures (i.e. galaxies not in groups, and not in pairs) to inhabit similar parameter spaces. These unpaired and ungrouped galaxies are blue, faint, and late type; but we do not see this for galaxies in voids. Curiously, group mass does not appear to correlate strongly with certain parameters: most notably S\'{e}rsic index, absolute magnitude, and ellipticity. Ungrouped galaxies are strongly unimodal in their colour distribution, suggesting a change in stellar populations driven by infall into groups. We see distributions for central and satellite galaxies in groups, as well as dominant and sub-dominant galaxies in pairs, that are consistent with other GAMA results from \citet{Robotham2013a}. The most visible trends are for galaxies classified by pair membership, in that galaxies in pairs show signs of being much more red (i.e. having suppressed star formation), and have a higher S\'{e}rsic index. These trends are seen again in Figure \ref{fig:ellipfrac}, which shows the fraction of galaxies that are early type to be much higher in pairs and higher mass groups (when one controls for stellar mass). 

Aware that mass matching to the void galaxy sample restricts our results to a narrow mass range, we generate a second set of mass controlled galaxies, this time matched to the GAMA GSMF from \citet{Baldry2012}. While galaxy properties continue to be unaffected by large scale structure for this higher mass matched sample, the distributions of certain galaxy types reflect the inclusion of higher masses. Central galaxies and dominant pair galaxies are now more likely to be late type, and all populations show a greater bimodality in their colour distributions; these results are consistent with the findings of \citet{Robotham2013a}. We further explore the effects of controlling for mass by removing it entirely for Figure \ref{fig:bigBinGrid}, where it is shown that galaxies in environments have a tendency to be correllated in similar ways as they go to higher stellar masses. The most massive galaxies in the Universe look the same regardless of where they are. Furthermore, removing mass normalisation also enhances trends in galaxy properties with structure as in Figure \ref{fig:sidewaysPlot_notNorm}, indicating strongly that stellar mass (or what drives stellar mass) is the dominant force in galaxy evolution. We see the most significant effects when looking at pair classifications, and is consistent with findings in \citet{Robotham2013a} and \citet{Robotham2014}. A macroscopic view of our results suggests that the properties of galaxies reviewed in this work correllate most strongly with stellar mass.

The results presented in this work serve as a broad overview of how large scale structure and environment directly impact on galaxy evolution, and serve as a broad overview of the trends of galaxies in different environments. These results motivate further study, in particular to detect mass flows within filaments by analysing the properties of spiral galaxies. This field stands to gain a much deeper understanding of the environmental mechanisms that influence galaxy evolution as more detailed surveys of gas in more distant galaxies become available with the advent of surveys such as ASKAP DINGO, followed by the SKA. The GAMA Large Scale Structure Catalogue presents a unique opportunity to study these processes within a well defined sample of galaxies in extremely underdense environments.

\section*{Acknowledgements}

The authors acknowledge the reviewer for their detailed and thoughtful comments on this work. MA is funded by an appointment to the NASA Postdoctoral Program at Ames Research Centre, administered by Oak Ridge Associated Universities through a contract with NASA; and acknowledges past funding from the University of Western Australia and the University of St Andrews. 

GAMA is a joint European-Australasian project based around a spectroscopic campaign using the Anglo-Australian Telescope. The GAMA input catalogue is based on data taken from the Sloan Digital Sky Survey and the UKIRT Infrared Deep Sky Survey. Complementary imaging of the GAMA regions is being obtained by a number of independent survey programs including GALEX MIS, VST KiDS, VISTA VIKING, WISE, Herschel-ATLAS, GMRT and ASKAP providing UV to radio coverage. GAMA is funded by the STFC (UK), the ARC (Australia), the AAO, and the participating institutions. The GAMA website is http://www.gama-survey.org/. The VISTA VIKING data used in this paper is based on observations made with ESO Telescopes at the La Silla Paranal Observatory under programme ID 179.A-2004.

\appendix

\footnotesize
\bibliographystyle{mn2e}
\setlength{\bibhang}{2.0em}
\setlength{\labelwidth}{0.0em}
\bibliography{galprops.bib}

\vspace{4mm}
\noindent \emph{$^1$NASA Ames Research Centre, N232, Moffett Field, Mountain View, CA 94035, United States\\
$^2$SUPA, School of Physics and Astronomy, University of St Andrews, North Haugh, St Andrews, Fife, KY16 9SS, UK\\
$^3$International Centre for Radio Astronomy Research, 7 Fairway, The University of Western Australia, Crawley, Perth,\\
 Western Australia 6009, Australia\\
$^4$Max Planck Institute fuer Kernphysik, Saupfercheckweg 1, 69117 Heidelberg, Germany\\
$^5$University of the Western Cape, Robert Sobukwe Road, Bellville, Cape Town 7535, South Africa\\
$^6$Institut f\"{u}r Astro- und Teilchenphysik, Universit\"{a}t Innsbruck, Technikerstra{\ss}e 25, 6020 Innsbruck, Austria\\
$^7$Department of Physics and Astronomy, Macquarie University, NSW 2109, Australia\\
$^8$Australian Astronomical Observatory, PO Box 915, North Ryde, NSW 1670, Australia\\
$^9$School of Physics, the University of Melbourne, Parkville, VIC 3010, Australia\\
$^{10}$School of Physics and Astronomy, The University of Nottingham, University Park, Nottingham, NG7 2RD, UK\\
$^{11}$Sydney Institute for Astronomy, School of Physics A28, University of Sydney, NSW 2006, Australia\\
$^{12}$School of Physics, Monash University, Clayton, Victoria 3800, Australia\\
$^{13}$Research School of Astronomy and Astrophysics, The Australian National University, Canberra, ACT 2611, Australia\\
$^{14}$Institute for Astronomy, University of Edinburgh, Royal Observatory, Blackford Hill, Edinburgh EH9 3HJ, UK\\
$^{15}$ESO, Karl-Swarzschild-Str. 2, D-85748 Garching bei M\"{u}nchen, Germany
$^{16}$Instituto de Astronom\'{i}a, Universidad Nacional Aut\'{o}noma de M\'{e}xico, A.P. 70-264, 04510 M\'{e}xico, D.F., M\'{e}xico\\
$^{17}$Astronomy Centre, University of Sussex, Falmer, Brighton BN1 9QH\\
$^{18}$Observatories of the Carnegie Institution for Science, 813 Santa Barbara Street, Pasadena, CA 91101, USA\\
$^{19}$Indian Institute of Science Education and Research Mohali- IISERM, Knowledge City, Sector 81, SAS Nagar, Manauli, PO 140306, India\\
$^{20}$Institute for Computational Cosmology, Department of Physics, Durham University, South Road, Durham, DH1 3LE, UK\\
$^{21}$Institute of Cosmology and Gravitation, University of Portsmouth, Burnaby Road, Portsmouth, PO1 3FX, United Kingdom\\
$^{22}$Department of Physics and Mathematics, University of Hull, Cottingham Road, Kingston-upon-Hull, HU6 7RX, UK\\
$^{23}$Jeremiah Horrocks Institute, University of Central Lancashire, PR1 2HE, Preston, UK\\
$^{24}$The Astronomical Institute of the Romanian Academy, Str. Cutitul de Argint 5, Bucharest, Romania}

\normalsize

\label{lastpage}

\end{document}